\documentclass[conference]{IEEEtran}
\usepackage{cite}
\usepackage{amsmath,amssymb,amsfonts}
\usepackage{algorithm}
\usepackage{algpseudocode}
\usepackage{graphicx}
\usepackage{textcomp}
\usepackage{marvosym}
\usepackage{xcolor}
\usepackage[hyphens]{url}
\usepackage{verbatim}

\usepackage{mathptmx} 
\usepackage{subfig}
\usepackage{color}
\usepackage{tikz}
\definecolor{yellow}{RGB}{255,217,101}      
\definecolor{blue}{RGB}{102,153,255}      
\definecolor{green}{RGB}{146,208,80}      

\def\BibTeX{{\rm B\kern-.05em{\sc i\kern-.025em b}\kern-.08em
    T\kern-.1667em\lower.7ex\hbox{E}\kern-.125emX}}

\usepackage{makecell}
\usepackage{multirow}
\usepackage{fancyhdr}
\usepackage{array}
\newcommand{\preserveBackslash}[1]{\let\temp=\\#1\let\\=\temp}
\newcolumntype{C}[1]{>{\preserveBackslash\centering}p{#1}}

\usepackage[colorlinks,
            linkcolor=blue,       
            anchorcolor=blue,  
            citecolor=blue,        
            ]{hyperref}

\title{SPLIM: Bridging the Gap Between Unstructured SpGEMM and Structured In-situ Computing}

\author{\normalsize{Huize Li, and Tulika Mitra\textsuperscript{\Letter}} \{huizeli, dcstm\}@nus.edu.sg}

\begin{document}
\maketitle
\thispagestyle{plain}
\pagestyle{plain}


\begin{abstract}

Sparse matrix-matrix multiplication (SpGEMM) is a critical kernel widely employed in machine learning and graph algorithms. However, real-world matrices' high sparsity makes SpGEMM memory-intensive. In-situ computing offers the potential to accelerate memory-intensive applications through high bandwidth and parallelism. Nevertheless, the irregular distribution of non-zeros renders SpGEMM a typical unstructured software. In contrast, in-situ computing platforms follow a fixed calculation manner, making them structured hardware. The mismatch between unstructured software and structured hardware leads to sub-optimal performance of current solutions.

In this paper, we propose SPLIM, a novel in-situ computing SpGEMM accelerator. SPLIM involves two innovations. First, we present a novel computation paradigm that converts SpGEMM into structured in-situ multiplication and unstructured accumulation. Second, we develop a unique coordinates alignment method utilizing in-situ search operations, effectively transforming unstructured accumulation into high parallel searching operations. Our experimental results demonstrate that SPLIM achieves 275.74$\times$ performance improvement and 687.19$\times$ energy saving compared to NVIDIA RTX A6000 GPU.

\end{abstract}

\section{Introduction}
\label{intro}

{\em Sparse matrix-matrix multiplication} (SpGEMM) is an essential operation for diverse applications, such as scientific computing~\cite{Azad2016, Ballard2016, Bell2012}, graph algorithms~\cite{Gilbert2011, Srivastava2020, Then2014}, and machine learning~\cite{Liu2015, Lu21, Qiu2022}. Real-world sparse matrices often exhibit large scale, lack structure, and high sparsity. The significant sparsity entails numerous ``zeros", substantially increasing SpGEMM's computational complexity. To mitigate computing complexity, researchers propose diverse compression methods to identify and skip unnecessary computation of zeros~\cite{Gilbert1992}. Moreover, the unstructured nature of sparse matrices leads to irregular distributions of non-zero values, resulting in substantial random memory access. In response to this issue, various software-oriented optimizations are proposed, such as compression formats~\cite{Compress2014, Gondimalla2019}, SpGEMM formulations~\cite{Akbudak2018, Demirci2020}, and load balancing strategies~\cite{Azad2016, Hypergraph2016}.

There is growing interest in hardware-oriented solutions, driven by the potential to improve execution efficiency through hardware-based algorithm optimization. Prominent architectures for accelerating SpGEMM include, GPU~\cite{Lee2020, Niu2022, Parger2020}, {\em Field Programmable Gate Array} (FPGA)~\cite{Haghi2020, Jamro2014, Lin2013}, {\em Application Specific Integrated Circuit} (ASIC)~\cite{gamma2021, Pal2018, Zhang2020}, and {\em Processing In Memory} (PIM)~\cite{Feng2022, Giannoula2022, Xie2021}. These solutions outperform software-oriented approaches by designing dedicated architectures and dataflows that efficiently support SpGEMM. Nonetheless, SpGEMM involves numerous random access in the whole memory space (as Figure~\ref{spmm} shows). Conventional hardwares suffer from substantial off-chip access overhead when searching the memory space~\cite{Xie2021}. PIM-based solutions reduce the off-chip access overhead by integrating {\em processing elements} (PEs) near the memory banks. However, on-chip PEs can efficiently access only their local memory banks, leading to longer access times for searching data stored in other banks, known as cross-bank transfers~\cite{Zhou2022}. Additionally, the use of on-chip PEs may reduce memory density and increase thermal concerns.

{\em Processing-using-memory} (PUM)~\cite{Chi16, Imani2019} performs in-situ computation, processing tasks directly in memory cells where the data is stored. Without the need of integrated on-chip PEs, PUM platforms present promising solutions for reducing cross-bank transfers and eliminating integration overhead. PUM-based solutions~\cite{Chi16, Imani2019, Shafiee16} demonstrate exceptional performance and energy efficiency, particularly in processing {\em general matrix multiplication} (GEMM), a typical structured kernel. However, using structured PUM platforms to accelerate unstructured SpGEMM may lead to sub-optimal performance. The {\em Sparse Matrix-vector Multiplication} (SpMV) computation paradigm employed by GraphR~\cite{Song18} serves as an example. GraphR involves three steps: sparse matrices $\to$ compression for storage $\to$ decompression for computing. Nevertheless, this paradigm does not fully exploit the potential of in-situ computing, as the decompression phase introduces matrix remapping with significant transmission overhead. Moreover, the decompressed matrix reintroduces zeros, reducing the utilization of in-situ computing hardware (as shown in Figure~\ref{unmatch}).

\begin{figure*}[t]
\centering
\includegraphics[width=16.5cm]{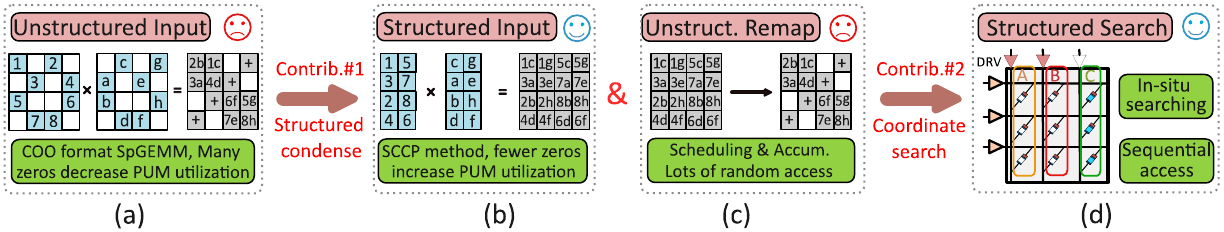}
\caption{(a) PUM-based SpGEMM using decompressed COO/CSR/CSC format, (b) Structured vector multiplication based on SCCP method, (c) Unstructured accumulation based on decompression, (d) Structured in-situ computing hardware}
\label{gap}
\vspace{-1em}
\end{figure*}

In response to the current landscape, we present \underline{SPLIM}\footnote{\underline{SPLIM}: unstructure \underline{Sp}GEMM \underline{l}inks to structure \underline{i}n-\underline{m}emory hardware}, a novel PUM-based SpGEMM accelerator. SPLIM introduces two innovations. First, we introduce a new computation paradigm for SpGEMM, namely {\em Structured Condensing Computing Paradigm} (SCCP). SCCP can perform in-situ structured vector multiplication with fewer zeros and higher hardware utilization, while the accumulation phase remains unstructured. Second, we adopt in-situ search operations for coordinates alignment, converting unstructured accumulation to high-parallel search operations. Figure~\ref{gap} offers a comprehensive overview of the insights behind SPLIM, depicting the transformation of unstructured SpGEMM into structured PUM-friendly kernels. In Figure~\ref{gap} (a), the COO/CSR/CSC computation paradigm used in PUM platforms is depicted, revealing an abundance of zeros that diminish hardware utilization. Figure~\ref{gap} (b) entails the adoption of structured vector multiplication utilizing the SCCP method, showing huge hardware utilization gains. Figure~\ref{gap} (c) showcases the accumulation of intermediate results through decompression, which introduces a multitude of on-chip scheduling and random access overhead. To address this, we introduce a search-based method for merging intermediate results, capitalizing on the in-situ computing capability of the PUM platform, as depicted in Figure~\ref{gap} (d). Our main contributions can be summarized as follows:

\begin{itemize}
    \item We analyze existing PUM-based SpGEMM accelerators, finding that they commonly use decompression to support unstructured SpGEMM. Decompression introduces many zeros to the PUM hardware, greatly reducing hardware utilization and system performance.
    \item We design SCCP, a novel SpGEMM computation paradigm that utilizes structured in-situ vector multiplication rather than decompression, increasing hardware utilization without introducing extra zeros.
    \item We perform coordinates alignment using in-situ search operations rather than decompression, effectively performing the unstructured accumulation without introducing scheduling overhead.
    \item We compare SPLIM with several modern SpGEMM accelerators. The experimental results show that SPLIM achieves superior performance and energy efficiency.
\end{itemize}


\section{Background and Motivation}
\label{backg}

\begin{figure}[t]
\centering
\includegraphics[width=8.7cm]{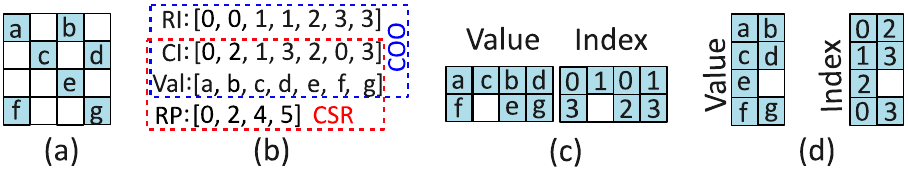}
\caption{(a) An example of sparse matrix, (b) The COO and CSR formats of the example, (c) The row-wise ELLPACK format, (d) The column-wise ELLPACK format}
\label{compress}
\vspace{-1em}
\end{figure}

\subsection{Sparse Matrix-matrix Multiplication}
Wilkinson defines sparse matrix as {\em any matrix with enough zeros that it pays to take advantage of them}~\cite{Gilbert1992}. Figure~\ref{compress} (a) illustrates an example of sparse matrix with white cells represent zeros, and blue cells represent non-zeros. To efficiently process sparse matrices, compression formats are employed for storage and processing, allowing the skipping of zeros and reducing computation complexity. Some widely-used compression formats include {\em coordinates format} (COO), {\em compressed sparse row} (CSR), {\em compressed sparse column} (CSC), {\em diagonal format} (DIA), and ELLPACK~\cite{ellpack1979}.

Figure~\ref{compress} (b) illustrates the COO format (blue dashed rectangle) of the example matrix, consisting of three vectors: the {\em row index} (RI), the {\em column index} (CI), and the {\em values} (Val). RI and CI store the row and column coordinates of Val, respectively. The CSR and CSC formats are modifications of the COO format, organized row-wise and column-wise, respectively. We present the CSR format in the red dashed rectangle, replacing the RI of COO format with a {\em row pointer} (RP) while CI and Val vectors unchanged. RP stores Val's index of the first non-zero element in each row, thus saving storage than RI. The DIA format performs well for sparse matrices with good diagonal locality. The ELLPACK format comprises two vectors: the Val vector stores non-zeros, and the index vector records the index of Val. As shown in Figure~\ref{compress} (c), the row-wise ELLPACK format condenses non-zeros to the topmost rows, while the index vector retains the original row index. Figure~\ref{compress} (d) presents the column-wise ELLPACK format, compressing non-zeros to the leftmost columns.

\begin{figure}[t]
\centering
\includegraphics[width=8.5cm]{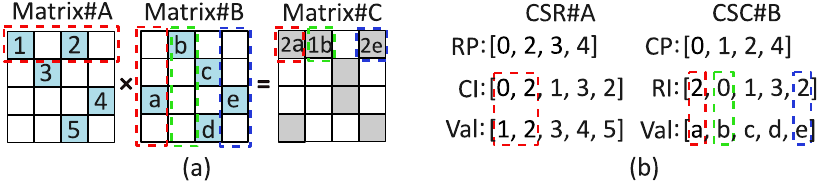}
\caption{(a) Matrix multiplication between sparse matrices $A$ and $B$, (b) Random access of CSR\#$A$ and CSC\#$B$}
\label{spmm}
\vspace{-1em}
\end{figure}

Figure~\ref{spmm} (a) depicts the SpGEMM operation $C = A \times B$. The three non-zeros' calculation in the first row of matrix $C$ is highlighted using red, green, and blue dashed rectangles. In Figure~\ref{spmm} (b), we present the CSR format of matrices $A$ and CSC format of matrix $B$, with the dashed rectangles representing the same meaning as in Figure~\ref{spmm} (a). However, due to the Val vector of the CSR and CSC formats do not contain index information for SpGEMM, we must realign their coordinates using the index vectors. Taking the example of red dashed rectangles in Figure~\ref{spmm} (b), only one multiplication will occur, as the coordinates alignment skips zeros. The results of coordinates alignment are shown by the three dashed rectangles in CSC\#B of Figure~\ref{spmm} (b), irregularly distributed in the entire Val vector. This irregular memory access of Val vectors renders SpGEMM a typical unstructured kernel. Since the COO, CSR, and CSC formats share the similar Val vectors, they have the same advantages and disadvantages when computing SpGEMM. This paper uses COOs referring to COO, CSR, and CSC formats.

\begin{figure}[t]
\centering
\includegraphics[width=7.5cm]{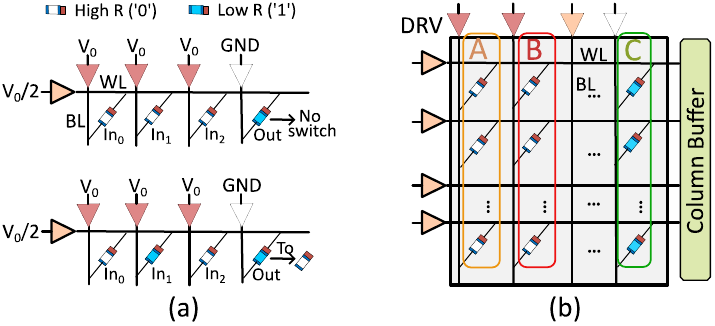}
\caption{(a) NOR operation of memristor switching, (b) Structured digital in-situ computing array}
\label{digital}
\vspace{-1em}
\end{figure}

\subsection{In-situ Computing}
\label{in-situ}
There are two types of in-situ computing: analog~\cite{Chi16} and digital in-situ computing~\cite{Imani2019}. Analog in-situ computing is less robust to noise due to the accumulation of analog signals. As SpGEMM is commonly used in neural networks and graph processing, requiring high precision, this paper focuses on the more noise-robust digital in-situ computing, which employs memristor switching to implement logic~\cite{Gupta18, Jang18, Siemon15}. In digital in-situ computing, a memristor cell has two states: high resistance (logic `0') and low resistance (logic `1'). By applying a proper voltage $V_0$, the memristor cell can switch from low resistance state `1' to high resistance state `0'. Researchers design digital in-situ computing with NOR logical gates~\cite{Siemon15} using above mechanisms.

Fig.~\ref{digital} (a) shows NOR operation $Out = NOR(In_0, In_1, In_2)$. The output memristor cell is initialized to `0', while the {\em word-line} (WL) is initialized to $\frac{V_0}{2}$. The {\em bit-lines} (BLs) of all input cells are activated with voltage $V_0$, while the bit-line of the output cell is linked to the {\em ground} (GND). When all the input cells ($In_0, In_1, In_2$) are in the high resistance state (`0'), no current flows from the BLs to the WL. Consequently, the voltage difference of the output cell remains at $\frac{V_0}{2}$, which is insufficient for state switching. On the other hand, if at least one input cell is in the low resistance state (`1'), the bit-line current flows from this cell to the word-line. As a result, the voltage difference of the output cell changes to $\frac{3V_0}{4}$, providing enough voltage for state switching from `1' to `0'.

Figure~\ref{digital} (b) presents the array-level digital in-situ computing. Initially, the output vector $C$ is initialized to `1'. The bit-lines of input vectors $A$ and $B$ are set to $V_0$, while the word-lines are set to $\frac{V_0}{2}$. By following the same procedure as shown in Figure~\ref{digital} (a), we can efficiently obtain $C = NOR(A, B)$ in a highly parallel manner. Due to the Turing completeness of NOR operation, we can perform numerous arithmetic/logic operations through a series of NOR operations~\cite{Imani2019}. Each memristor array is equipped with a column buffer, which is equivalent to DRAM's row buffer. The logical cells of a memristor array are coupled in a row and column manner, creating a structured architecture.

\subsection{Motivation}
\label{motivation}
{\bf PIM vs. PUM.} PIM platforms~\cite{Zhou2022} integrate PEs into memory, enabling in-memory computing that reduces off-chip transfers. However, due to area and thermal constraints, PIM platforms can only accommodate a limited number of logic units, which restricts computing parallelism. Moreover, significant data transfers occur between the on-chip logic units and memory banks, leading to conflicts in the {\em control and address} (C/A) buses shared by all memory banks. These conflicts arise when PIM platforms process large scale sparse matrix multiplication~\cite{Xie2021}. On the other hand, PUM platforms~\cite{Imani2019} show promise in reducing integration and transmission overhead. As depicted in Figure~\ref{digital} (b), each memristor cell in PUM platforms can function as a logic cell without integrating on-chip PEs, exposing million-level row parallelism. Furthermore, PUM platforms perform calculations directly in where the data is stored, thus reducing data transmission between storage and computation units. In comparison to PIM platforms, PUM platforms offer higher computational parallelism, lower integration complexity, and reduced transmission overhead.


\begin{figure}[t]
\centering
\includegraphics[width=8cm]{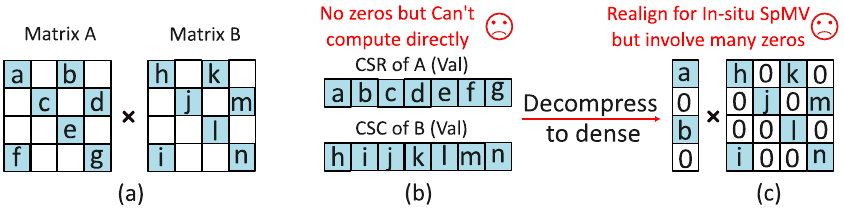}
\caption{(a) SpGEMM between matrices $A$ and $B$, (b) The Val of CSR\#$A$ and CSC\#$B$, (c) In-situ SpMV in GraphR~\cite{Song18}}
\label{unmatch}
\vspace{-1em}
\end{figure}

{\bf PUM + SpGEMM.} The strength of PUM platforms lie in their dedicated hardware designs, achieved at the expense of hardware flexibility. Figure~\ref{digital} (b) illustrates that all memristor cells are arranged in a regular row/column manner, providing million-level row parallelism, i.e., {\em the latency of processing one row is the same as processing millions of rows}, which efficiently handles structured GEMM. To utilize the high parallelism of PUM platforms, GraphR~\cite{Song18} proposes in-situ {\em sparse matrix-vector multiplication} (SpMV) with COOs format, which is widely used in PUM platforms~\cite{Huang2022, Lyu2023,Challapalle20}. Performing SpGEMM in Figure~\ref{unmatch} (a) involves multiple iterations of GraphR's SpMV kernel. As Figure~\ref{unmatch} (b) shows, the compressed Val vector of the COOs format cannot be directly used for SpGEMM due to unmatched coordinates. Therefore, GraphR decompresses the Val vector back to the structured dense matrix, reintroducing many zeros. Although PUM platform has millions-level row parallelism, if too many zeros are involved, the number of valid computations decreases (just like Figure~\ref{unmatch} (c)). In Figure~\ref{unmatch} (c), the white cells (zeros) waste calculation resources, referring to as invalid rows.

{\bf Our goal:} PUM platforms offer excellent hardware performance, holding promise for high parallelism in accelerating SpGEMM. Nevertheless, current PUM-based SpGEMM accelerators rely on decompression to process unstructured COOs format. The decompression approach reintroduces zeros, substantially reducing PUM utilization. To this end, we aim to design a more efficient method to bridge the gap between unstructured SpGEMM and structured PUM platforms.

\section{Structured In-situ SpGEMM}
\label{paradigm}

\subsection{Structured Condensing Computation Paradigm}
\label{disad}
{\bf High-level motivation.} To perform SpGEMM with PUM platforms, the Val vector of COO/CSR/CSC format must be realigned with index vectors. Decompression is the commonly method to align Val vectors in PUM platforms, as depicted in Figure~\ref{unmatch} (c). However, the decompression phase reintroduces zeros and hampers the utilization of PUM platforms. In Figure~\ref{unmatch} (c), the white cells (zeros) waste calculation resources, referring to as invalid rows, i.e., {\em more zeros in PUM} $\to$ {\em less valid computing} $\to$ {\em lower utilization of PUM}. We seek a potential solution to tackle the above issues by adopting a novel method, which can perform SpGEMM without decompressing the Val vector, i.e., computing directly using the compressed Val vector without decompression to introduce zeros. To this end, the number of valid calculations in PUM can be greatly increased without introducing extra zeros.

\begin{figure}[t]
\centering
\includegraphics[width=8.5cm]{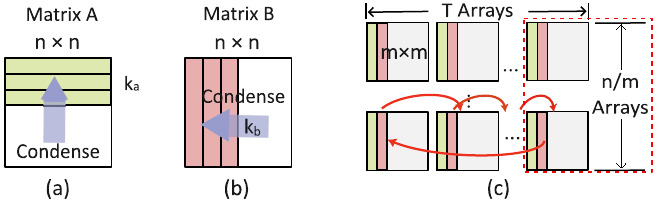}
\caption{(a) ELLPACK format of matrix $A$, (b) ELLPACK format of matrix $B$, (c) Matrices mapping of A and B}
\label{partition}
\vspace{-1em}
\end{figure}

{\bf Opportunities from ELLPACK format.} As depicted in Figure~\ref{unmatch}, the primary objective of the decompression operation is to align the coordinates of the two input Val vectors. {\em If the coordinates of two input Val vectors are naturally aligned, then we can eliminate the decompression for coordinates alignment. We identify that the SpGEMM coordinates alignment can be effectively divided into two distinct parts.} First, the {\em vector multiplication} only requires aligning the column coordinates of the left matrix with the row coordinates of the right matrix. Second, the {\em accumulation phase} merges the intermediate results generated by the vector multiplication, necessitating alignment of the row coordinates of the left matrix with the column coordinates of the right matrix. This observation presents an opportunity for computing {\em vector multiplication} directly using the compressed Val vectors. Specifically, the row-wise ELLPACK format retains the column coordinates of left Val, while the column-wise ELLPACK format preserves the row coordinates of right Val. Consequently, the ELLPACK's Val vector of input matrices can be directly used to perform in-situ {\em vector multiplication} without the need for decompression.

{\bf Matrices mapping strategy.} In SpGEMM, processing matrices comprising millions of rows/columns needs to store the sparse matrices into multiple memristor arrays due to one array's limited capacity. Figure~\ref{partition} (a) presents the $n\times n$ sparse matrix $A$, where all non-zeros are condensed to the top, resulting in an ELLPACK format with $k_a$ row vectors. Similarly, Figure~\ref{partition} (b) shows the $n\times n$ sparse matrix $B$, with non-zeros condensed to the left, yielding an ELLPACK format with $k_b$ column vectors. To map the ELLPACK format of matrices $A$ and $B$, we utilize multiple $m\times m$ memristor arrays, as presented in Figure~\ref{partition} (c). Specifically, each memristor array stores a row vector of $A$ and a column vector of $B$. Since $n\gg m$, $\frac{n}{m}$ memristor arrays are used to store the same vector (red dotted rectangle). To accommodate $k_a$ row vectors of matrix $A$ and $k_b$ column vectors of matrix $B$, we employ $T$ memristor arrays. Each memristor array is responsible for storing $\frac{k_a}{T}$ row vectors of matrix $A$ and $\frac{k_b}{T}$ column vectors of matrix $B$. When $k_a = k_b = T$, the storage configuration of matrices $A$ and $B$ is depicted in Figure~\ref{partition} (c).

\begin{figure}[t]
\centering
\includegraphics[width=8.5cm]{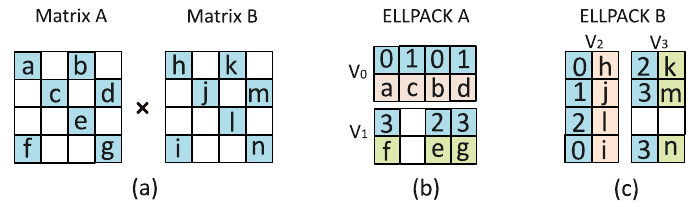}
\caption{(a) An example of SpGEMM $A\times B$, (b) ELLPACK format of matrix $A$, (c) ELLPACK format of matrix $B$}
\label{examp}
\vspace{-1em}
\end{figure}

{\bf ELLPACK-based in-situ vector multiplication.} We employ matrices $A$ and $B$ shown in Figure~\ref{examp} (a) as an example. Figure~\ref{examp} (b) presents the row-wise ELLPACK format of matrix $A$ with $k_a = 2$ row vectors, namely $V_0$ and $V_1$. Similarly, Figure~\ref{examp} (c) shows the column-wise ELLPACK format of matrix $B$ comprising $k_b = 2$ column vectors, denoted as $V_2$ and $V_3$. Each row/column vector is composed of two parts: the index vector (numbers) and the Val vector (alphabets).

\begin{figure}[t]
\centering
\includegraphics[width=8.5cm]{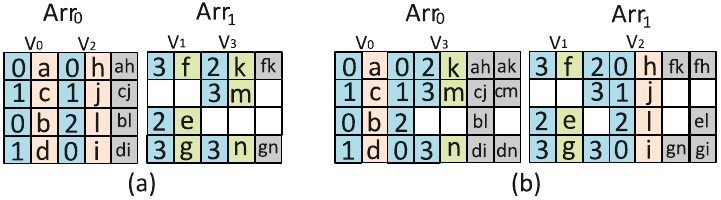}
\caption{(a) The data mapping of matrices $A$ and $B$, (b) The intermediate results of in-situ vector multiplication}
\label{vm}
\vspace{-1em}
\end{figure}

To map the ELLPACK format of matrices $A$ and $B$, we follow the matrix mapping approach depicted in Figure~\ref{partition} (c). In our example, we use two memristor arrays, namely Arr$_0$ and Arr$_1$, to store the row vectors of $A$ and the column vectors of $B$. Specifically, Arr$_0$ stores $V_0$ and $V_2$, while Arr$_1$ stores $V_1$ and $V_3$. Arr$_0$ and Arr$_1$ will conduct the in-situ vector-vector multiplication directly using the Val vector of ELLPACK formats (no decompression for coordinates alignment), leading to the generation of intermediate results (indicated by grey columns) shown in Figure~\ref{vm} (a).

The SpGEMM kernel involves iterative vector multiplication among different vectors. However, storing sparse matrices in separate memristor arrays necessitates cross-array data transfers, leading to increased random access. To mitigate this issue, we employ the ring-wise broadcast method, illustrated by the red arrows in Figure~\ref{partition} (c). Specifically, each column vector of matrix $B$ is transferred to the next memristor array, i.e., $Arr_i \to Arr_{i+1}$. In our example, we transfer $V_1$ to Arr$_1$ and $V_3$ to Arr$_0$. Arr$_0$ and Arr$_1$ then perform the in-situ vector-vector multiplication again, generating the next intermediate results indicated by the grey columns in Figure~\ref{vm} (b). To optimize memory usage, we overwrite the previous Val vector with the newly received one. However, since all the index vectors are essential for coordinate alignment, we retain the previous index vector instead of rewriting it.

{\bf Latency analysis.} Figure~\ref{unmatch} shows the in-situ SpMV using the decompressed COOs format, which requires $4\times$ SpMV iterations for SpGEMM. Figure~\ref{vm} depicts our ELLPACK-based computation paradigm, achieving the same vector multiplication in only $2\times$ iterations. Compared to the COOs computation paradigm, our ELLPACK computation paradigm shows 2$\times$ iterations saving due to reduce 50\% zeros. In realistic sparse matrices with $<1\%$ sparsity (over 99\% are zeros), our ELLPACK-based computation paradigm can save over 99$\times$ iterations (latency) by eliminating zeros.

{\bf Transmission analysis.} We adopt ring-wise broadcast to perform the vector multiplication among Val vectors stored in different arrays, totally $T\times$ ring-wise broadcast for $T\times$ arrays. Fortunately, the cross-array transmission is more efficient than cross-bank transmission. Pinatubo~\cite{Pinatubo2016} shows that we can perform high performance RowClone~\cite{rowclone2013} between different memristor arrays using the column buffer. Specifically, we numbered the memristor arrays as odd ($2i-1$) and even ($2i$), $i = 1, 2, ..., \frac{T}{2}$. We can finish one ring-broadcast with two steps RowClone. {\em In the first RowClone}, we read arrays $2i-1$ to their column buffer. Then, we transfer the data from column buffer $2i-1$ to $2i$. Finally, we write data from column buffer to array $2i$. {\em In the second RowClone}, we perform the above operation from $2i$ to $2i+1$. Therefore, we need only $2T\times$ RowClone to finish all ring-broadcast without C/A bus conflicts.

{\bf Memory analysis.} SCCP performs $T\times$ iterations of vector multiplication to generate an intermediate vector in each iteration, totally $T\times$ intermediate vectors for each memristor array. Thus, one memristor array is unable to store all these intermediate results. To scale memristor array, we adopt the {\em Block Size Scalability} (BSS) in FloatPIM~\cite{Imani2019}. For example, 32 $1k\times 1k$ arrays can form a $1k\times 32k$ array with BSS method.



\begin{figure}[t]
\centering
\includegraphics[width=8cm]{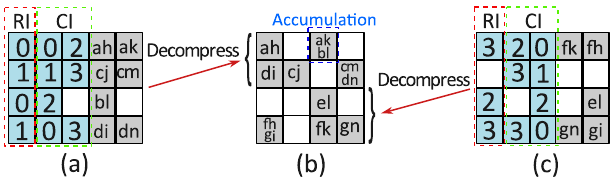}
\caption{(a) Intermediate results generated by Arr$_0$, (b) The decompressed intermediate results and accumulation operation, (c) Intermediate results generated by Arr$_1$}
\label{decom}
\vspace{-1em}
\end{figure}

\subsection{Accumulation Based on In-situ Search}
\label{accumulation}
{\bf High-level motivation.} The proposed ELLPACK-based computation paradigm exhibits high PUM utilization when processing in-situ vector multiplication. However, accumulating the intermediate results generated by this paradigm requires huge scheduling overhead, because the Val of ELLPACK format does not hold the row/column index of left/right matrix. Therefore, we still need to restore the coordinates using the index vectors before performing accumulation. Decompression is commonly employed to restore coordinates and accumulate intermediate results. Figure~\ref{decom} (a) and (c) depicts the intermediate results from Section~\ref{disad}, where the {\em row index} (RI) and {\em column index} (CI) are marked with red and green dashed rectangles, respectively. Figure~\ref{decom} (b) shows the SpGEMM output matrix obtained through accumulating the decompressed intermediate results with RI and CI.

However, decompression-based methods introduce two challenges: substantial scheduling overhead and storage overhead. First, the decompression operation typically relies on the on-chip scheduler, which utilizes RI and CI of the intermediate results to generate control signals for decompression. Given the large number of randomly distributed intermediate results, the on-chip scheduling overhead can become substantial. Second, the decompression operation converts the ELLPACK format of intermediate results to a decompressed dense matrix, imposing considerable storage overhead for zeros. These challenges underscore the need for more efficient methods that can mitigate the scheduling and storage overheads.

\begin{figure}[t]
\centering
\includegraphics[width=8.5cm]{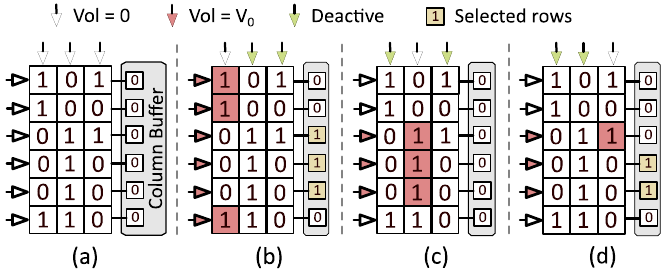}
\caption{(a) Int3 unsigned vector for in-situ search operation, (b) DRVs status of bit-2, (c) DRVs status of bit-1, (d) DRVs status of bit-0}
\label{min}
\vspace{-1em}
\end{figure}

{\bf Opportunities from in-situ search.} We find that the primary objective of decompression is to extract intermediate results with identical row and column coordinates, as depicted in Figure~\ref{decom}. A novel approach that extracts all intermediate results with matching coordinates, without relying on a scheduler or data remapping, could significantly reduce the scheduling and storage overhead. In this context, the in-situ search operation of the memristor array emerges as a promising solution for high parallel in-situ coordinates alignment.

{\bf Introduction of in-situ search.} The in-situ search operation of memristor arrays is illustrated in Figure~\ref{min}. For visualization, we use six 3-bit unsigned integers as an example, shown in Figure~\ref{min} (a) (Alg.~\ref{alg:min} line 1). Three distinct colors are used to indicate the voltage driver status: white for $V=0$, red for $V=V_0$, and green for deactivated status. During each iteration, the column buffer is set to all zeros (Alg.~\ref{alg:min} line 5). The in-situ search starts by setting all row DRVs to $V_0$ (Alg.~\ref{alg:min} line 3), resulting in high voltage word-lines, as depicted in Figure~\ref{min} (b). Next, we deactivate all column DRVs except the highest bit (bit-2) while setting its DRV to zero (Alg.~\ref{alg:min} line 6). The voltage difference enables the word-lines' current flowing from the low resistance (`1') cells to the bit-line, represented by the red cells in Figure~\ref{min} (b). Meanwhile, the high resistance (`0') cells prevent current flow, maintaining high voltage on the word-lines, which is recorded by the column buffer (yellow cells and Alg.~\ref{alg:min} line 7).

The above process is iterated for the next bit. As shown in Figure~\ref{min} (c), the row DRVs are set based on the `1' signals stored in the column buffer (Alg.~\ref{alg:min} line 8). We deactivate all column drivers except for bit-1 while setting its DRV to zero. All memristor cells are in low resistance (`1') states, so all word-lines' current flow to the bit-line. {\em Consequently, the column buffer records no high voltage (`1'), and the same row DRVs are activated in the subsequent iteration (bit-0).} Figure~\ref{min} (d) provides details of the lowest bit (bit-0) processing. Following the same principle of leakage current, the column buffer records word-lines with high voltage, representing the storage of minimal value. 

\begin{algorithm}
\caption{In-situ Minima Search}\label{alg:min}
\begin{algorithmic}[1]
\Require Unsigned 32-bit integer vector $V$ $\in$ $\mathbb{R}^{n}$.
\Ensure Minimum value of $V$.
\State Mapping vector $V$ to $n$-rows ReRAM (Figure~\ref{min} (a)).
\State Initializing $i$ to highest bit with $i=32$.
\State Activating all row DRVs with $V = V_0$ (Figure~\ref{min} (b)).
\While{$i > 0$} 
\State Initializing column buffer to `0'.
\State Activating $i$-th column DRV with $V = 0$ (Fig.~\ref{min}(b)).
\State Column buffer (CB) stores `1' signals (Figure~\ref{min} (b)).
\State Activating DRVs of row's CB stores `1' (Fig.~\ref{min} (c)). \Comment{If no row's CB stores `1', row DRVs' activation remains the same as previous iteration (Figure~\ref{min} (d))}
\State $i=i-1$.
\EndWhile
\State Return all rows' CB stores `1' (Figure~\ref{min} (d)).
\end{algorithmic}
\end{algorithm}

To find the next minimal value, we need to deactivate the rows storing the current minimal value. Repeating the above iterations allows us to identify those rows holding the subsequent minimal values. ReSQM~\cite{resqm2020} introduces an in-situ sorting algorithm for structured database queries. Their method draws inspiration from the bit-wise minimal algorithm for {\em Resistive Content Addressable Memory} (ReCAM). However, it is worth noting that ReSQM's method requires double hardware resources compared to our approach due to the fact that one ReCAM cell is composed of two memristor cells.

\begin{figure}[t]
\centering
\includegraphics[width=8.1cm]{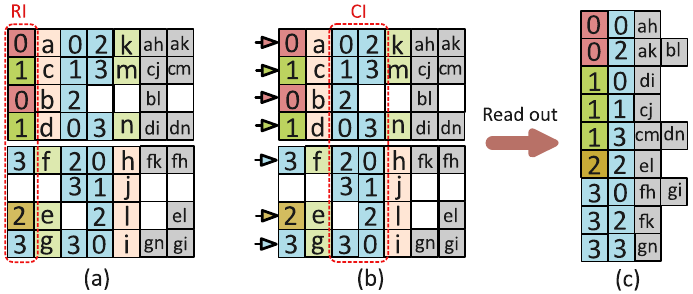}
\caption{(a) In-situ search operation for RI, (b) In-situ search operation for CI, (c) Sorted list of output matrix}
\label{search}
\vspace{-1em}
\end{figure}

{\bf Detailed designs.} In Figure~\ref{search} (a), we present the intermediate results obtained from Section~\ref{disad}. The {\em row index} (RI) of matrix $A$ is indicated by the red dotted rectangle. First, we perform an in-situ search operation on the RI to locate all rows with the smallest row index (RI\#0 marked in red cells). Subsequently, we activate only these rows with RI\#0 (red DRVs in Figure~\ref{search} (b)) and conduct another in-situ search operation on the {\em column index} (CI). In our example, the smallest CI is CI\#0. Consequently, we retrieve the intermediate results of RI\#0 and CI\#0, i.e., `ah' in Figure~\ref{search} (c). We then store RI\#0, CI\#0, and `ah' to the off-chip memory. We modify the sign bit of CI\#0 to `1', rendering this CI as invalid. Next, we iterate to search for the next minimal CI of RI\#0, which in our example is CI\#2. Then, we add the intermediate results of RI\#0 and CI\#2, i.e., `ak' and `bl' using an on-chip accumulator to obtain the Val of RI\#0 and CI\#2.

Once we complete the CI iterations of RI\#0, we set the sign bit of RI\#0 to `1', making RI\#0 invalid. We proceed to search for the next minimal RI, which is RI\#1 in our example (green cells in Figure~\ref{search} (a)). We then activate rows of RI\#1 (green DRVs in Figure~\ref{search} (b)) and conduct an in-situ search on the CI of RI\#1. Following the same procedure to all RI, we can obtain a sorted list as Figure~\ref{search} (c) shows. After each row's iteration, we add all grey cells using an on-chip accumulator to obtain the COO format of the output matrix.

{\bf Memory/Transmission/Latency analysis.} {\em Memory}: Apart from storing the output COO format in off-chip memory, we do not need additional memory, thanks to the use of in-situ search rather than decompression. {\em Transmission}: After the in-situ search, we can read data with exact access. Therefore, only one iteration of full memory space exact access can obtain the final results. {\em Latency}: For two $n\times k$ ELLPACK matrix, we can obtain all rows of the output matrix with only $n\times$ RI iterations. In each RI iteration, we need $k\times$ CI iterations, resulting in an $\mathbf{O}(nk)$ complexity. For highly sparse matrices, $n\gg k$, and we have $\mathbf{O}(n)$ time complexity. In comparison to the decompression-based method, our approach eliminates the need for an on-chip scheduler. Additionally, our method allows us to generate the COO format of the output matrix without requiring decompression, thereby reducing the significant random access and on-chip storage overhead.

\subsection{Hybrid ELLPACK and COO Format}
\label{optimization}
\begin{figure}[t]
\centering
\includegraphics[width=8.1cm]{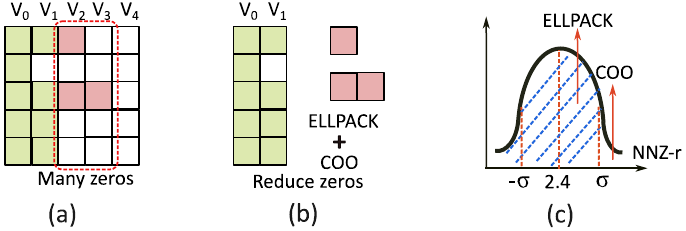}
\caption{(a) ELLPACK format with zeros, (b) Hybrid ELLPCAK and COO formats~\cite{Bernabeu2015}, (c) Normal distribution of NNZ-r}
\label{hybrid}
\vspace{-1em}
\end{figure}

The ELLPACK format~\cite{ellpack1979} is well-suited for sparse matrices with a random distribution, wherein all non-zero values are randomly and uniformly distributed. However, real-world matrices often do not exhibit complete randomness and may contain rows or columns with good locality. In Figure~\ref{hybrid} (a), the column-wise ELLPACK format of a sparse matrix comprises four column vectors labeled from $V_0$ to $V_3$. We use two colors to illustrate the impact of non-zero values' distributions. The green color vectors, like $V_0$ and $V_1$, demonstrate a high compression ratio with only a few zeros. On the other hand, the red color vectors, like $V_2$ and $V_3$, exhibit a low compression ratio with many zeros. Thus, when compressing sparse matrices using the ELLPACK format, the red vectors may yield a suboptimal compression ratio. To address this limitation, as shown in Figure~\ref{hybrid} (b), we utilize the COO format to store the red vectors, providing a hybrid approach~\cite{Bernabeu2015} that improves overall compression efficiency.

We employ three metrics, NNZ-r, NNZ-a, and $\sigma$, to establish the boundary between the ELLPACK format and the COO format. NNZ-r represents the number of non-zeros in each row, NNZ-a denotes the average non-zeros across all rows, and $\sigma$ represents the standard deviation of NNZ-r. In Figure~\ref{hybrid} (c), we illustrate the normal distribution of NNZ-r. Rows with NNZ-r $\leq$ NNZ-a + $\sigma$ are stored using the ELLPACK format, while rows with NNZ-r $>$ NNZ-a + $\sigma$ are stored using a hybrid combination of the ELLPACK and COO formats.

\section{SPLIM}
\label{archi}

\subsection{Overview Architecture}
\begin{figure}[t]
\centering
\includegraphics[width=8.5cm]{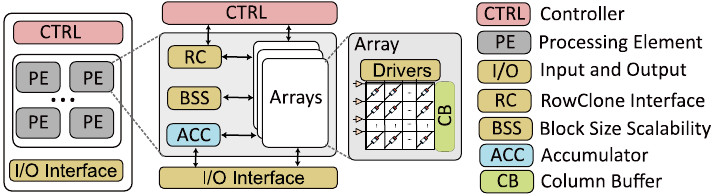}
\caption{SPLIM architecture}
\label{splim}
\vspace{-1em}
\end{figure}

Figure~\ref{splim} illustrates the architecture of SPLIM, comprising a group of {\em processing elements} (PEs), a {\em controller} (CTRL), and {\em input and output} (I/O) interfaces. Each PE houses {\em RowClone interfaces} (RC), {\em block size scalability} (BSS), {\em Accumulators} (ACC), and multiple memristor arrays. The functions of RC (for ring-wise broadcast), BSS (scale array size), and ACC are same as described in Section~\ref{paradigm}. The memristor arrays consist of numerous row DRVs and column DRVs. The CTRL sends identical control signals to the row/column DRVs of all PEs, enabling parallel processing across the memristor arrays. Each memristor array operates in two states: memory status and computation status. In memory status, the data for read/write operations are stored in the {\em column buffer} (CB). In computation status, PEs will perform in-situ calculations introduced in Section~\ref{paradigm} by applying appropriate voltages.



\subsection{SPLIM Dataflow}
\label{dataflow}
To present the end-to-end SpGEMM dataflow, we assume that the two input matrices are pre-stored in the PEs of SPLIM. We use the hybrid ELLPACK and COO formats in Section~\ref{optimization} to store input matrices. We segregate the storage of ELLPACK format and COO format in different PEs. Those storing ELLPACK format are referred to as ELL-PEs, while those storing COO format are called COO-PEs.

Initially, all ELL-PEs perform the in-situ vector multiplication, generating and storing intermediate results. Subsequently, each ELL-PE transmits the vector of the right matrix to its neighboring ELL-PE, following the two steps RowClone illustrated in Figure~\ref{partition} (c). The process is then repeated, with all ELL-PEs performing the in-situ vector multiplication again to generate further intermediate results. The above RowClone operation and in-situ vector multiplication iterate until all intermediate results are obtained. Finally, all PEs execute the in-situ search operation to serially send the COO format of the output matrix to the accumulator for merging, the result of which is stored to the off-chip memory.

On the other hand, the COO-PEs will process these good locality rows stored with the COO format. All ELL-PEs are working in memory status, and the COO-PEs can access data from the ELL-PEs. This type of random memory access does not increase on-chip bandwidth pressure, because the involved rows have good row locality. The calculation of the COO format part follows the procedure introduced in Figure~\ref{unmatch}.

\subsection{Comparison of Memory and Time Complexity}
\label{complexity}
To quantitatively analyze the superiority of the proposed computation paradigm, we configure COO/CSR/CSC-SPLIM (short for COO-SPLIM), implementing the COOs computation paradigm shown in Figure~\ref{unmatch} on SPLIM. For the sake of brevity, we assume that both the left and right matrices are randomly distributed $N \times N$ sparse matrices, with a standard deviation of $\sigma = 0$. Additionally, for both the left and right matrices, we assume that their NNZ-a = NNZ-r = $K$, indicating that the ELLPACK format comprises $K$ vectors.

\begin{table*}[pbt]
\centering
\tabcolsep=0.12cm
    \caption{Matrices from SuiteSparse Matrix Collection~\cite{Davis2011}, {\bf Dim.} is the number of rows/columns and $nnz_{av}$ is the average number of non-zero values per row, $\sigma$ is the stardard deviation of $nnz_{av}$.}
    \small
    \renewcommand\arraystretch{1}
    \label{tab:matrices}
    \tabcolsep=0.05cm
    \begin{tabular}{|c|c|c|c|c||c|c|c|c|c||c|c|c|c|c||c|c|c|c|c|}  
    \cline{1-20}
    
       {\bf ID}&{\bf Matrix} & {\bf Plot} & {\bf \makecell{Dim.\\$nnz$}} &{\makecell{$nnz_{av}$ \\$\sigma$}} & {\bf ID} & {\bf Matrix} & {\bf Plot} & {\bf \makecell{Dim.\\$nnz$}} &{\makecell{$nnz_{av}$ \\$\sigma$}} & {\bf ID}& {\bf Matrix} & {\bf Plot} & {\bf \makecell{Dim.\\$nnz$}} &{\makecell{$nnz_{av}$ \\$\sigma$}} & {\bf ID} & {\bf Matrix} & {\bf Plot} & {\bf \makecell{Dim.\\$nnz$}} &{\makecell{$nnz_{av}$ \\$\sigma$}}\\
       \hline \hline
       {\#1}&{\makecell{pdb1\\HYS}} & \begin{minipage}[s]{0.08\columnwidth}\raisebox{-.5\height}{\includegraphics[width=\linewidth]{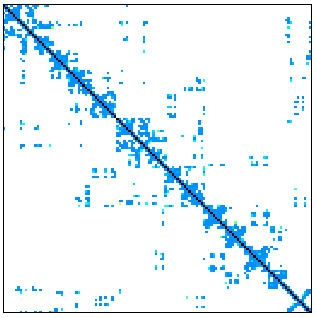}}\end{minipage} & {\makecell{36K\\4.3M}} &{\makecell{119.3\\31.86}}& {\#2} & {rma10} & \begin{minipage}[s]{0.08\columnwidth}\raisebox{-.5\height}{\includegraphics[width=\linewidth]{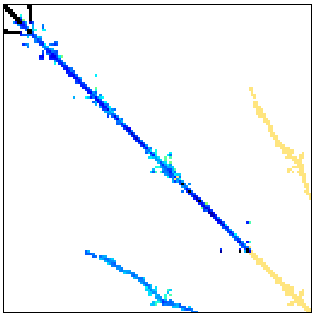}}\end{minipage} & {\makecell{47K\\2.3M}} & {\makecell{49.7\\27.78}}& {\#3} & {\makecell{bcss\\tk32}} & \begin{minipage}[s]{0.08\columnwidth}\raisebox{-.5\height}{\includegraphics[width=\linewidth]{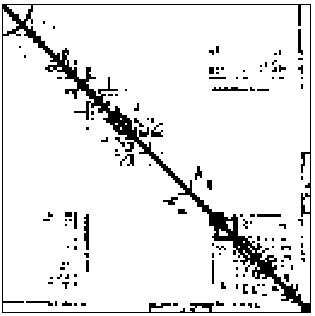}}\end{minipage} & {\makecell{45K\\2M}} & {\makecell{45.2\\15.48}}& {\#4} & {\makecell{ct20\\stif}} & \begin{minipage}[s]{0.08\columnwidth}\raisebox{-.5\height}{\includegraphics[width=\linewidth]{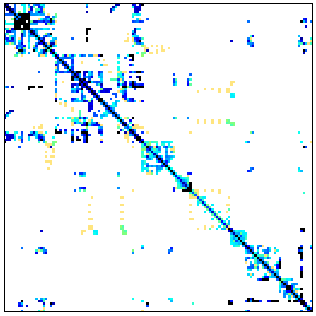}}\end{minipage} & {\makecell{52K\\2.6M}} &{\makecell{49.7\\16.98}}  \\
       \hline
       {\#5}&{cant} & \begin{minipage}[s]{0.08\columnwidth}\raisebox{-.5\height}{\includegraphics[width=\linewidth]{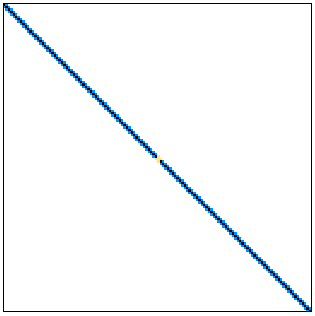}}\end{minipage} & {\makecell{62K\\4M}} & {\makecell{64.2\\14.06}} & {\#6} & {\makecell{crank\\seg\_2}} & \begin{minipage}[s]{0.08\columnwidth}\raisebox{-.5\height}{\includegraphics[width=\linewidth]{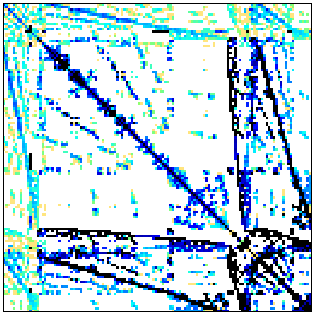}}\end{minipage} & {\makecell{64K\\14M}} & {\makecell{222\\95.88}} & {\#7} & {lhr71} & \begin{minipage}[s]{0.08\columnwidth}\raisebox{-.5\height}{\includegraphics[width=\linewidth]{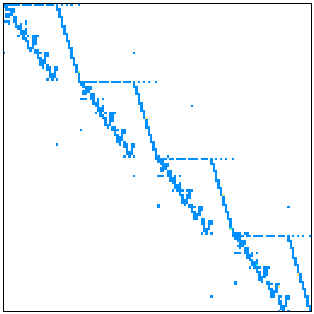}}\end{minipage} & {\makecell{70K\\1.5M}} & {\makecell{21.3\\26.32}} & {\#8} & {consph} & \begin{minipage}[s]{0.08\columnwidth}\raisebox{-.5\height}{\includegraphics[width=\linewidth]{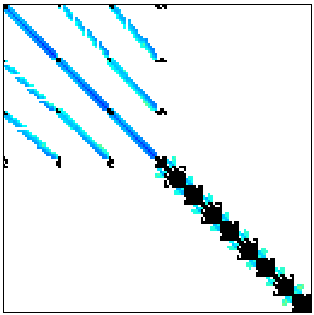}}\end{minipage} & {\makecell{83K\\6M}} & {\makecell{72.1\\19.08}}  \\
       \hline
       {\#9}&{\makecell{soc-sign-\\epinions}} & \begin{minipage}[s]{0.08\columnwidth}\raisebox{-.5\height}{\includegraphics[width=\linewidth]{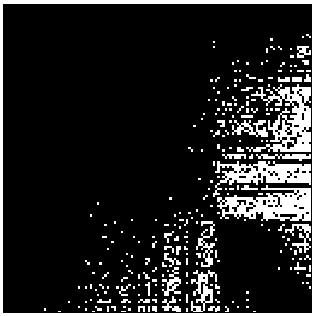}}\end{minipage} & {\makecell{132K\\841K}} & {\makecell{6.4\\32.95}} & {\#10} & {\makecell{ship\\sec1}} & \begin{minipage}[s]{0.08\columnwidth}\raisebox{-.5\height}{\includegraphics[width=\linewidth]{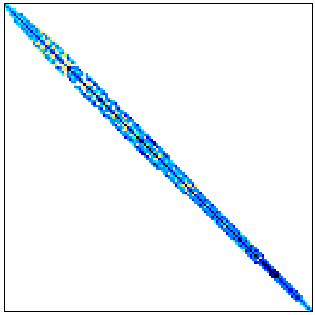}}\end{minipage} & {\makecell{141K\\3.6M}} & {\makecell{25.3\\11.07}} &{\#11} & {xenon2} & \begin{minipage}[s]{0.08\columnwidth}\raisebox{-.5\height}{\includegraphics[width=\linewidth]{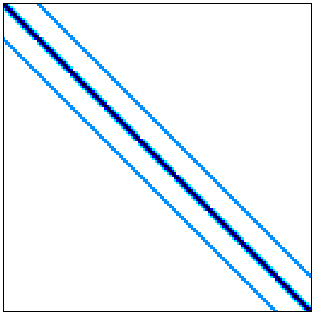}}\end{minipage} & {\makecell{157K\\3.9M}} & {\makecell{24.6\\4.07}} & {\#12} & {ohne2} & \begin{minipage}[s]{0.08\columnwidth}\raisebox{-.5\height}{\includegraphics[width=\linewidth]{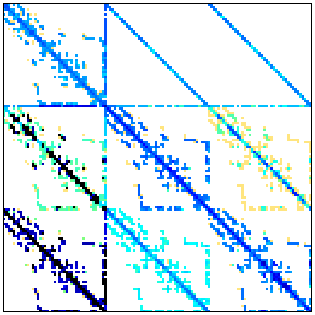}}\end{minipage} & {\makecell{181K\\6.9M}} & {\makecell{37.9\\21.09}} \\
       \hline
       {\#13}&{pwtk} & \begin{minipage}[s]{0.08\columnwidth}\raisebox{-.5\height}{\includegraphics[width=\linewidth]{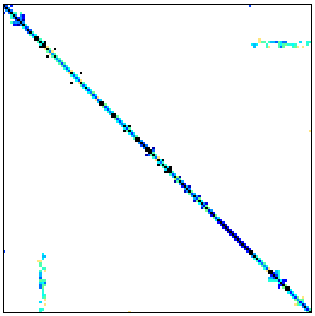}}\end{minipage} & {\makecell{218K\\11.5M}} & {\makecell{52.9\\4.74}} & {\#14} & {\makecell{stan\\ford}} & \begin{minipage}[s]{0.08\columnwidth}\raisebox{-.5\height}{\includegraphics[width=\linewidth]{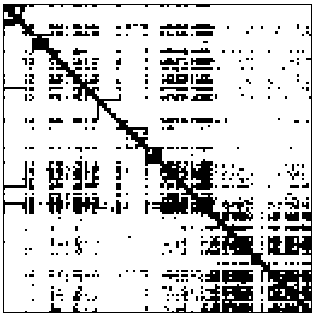}}\end{minipage} & {\makecell{282K\\2.3M}} & {\makecell{8.2\\166.33}} & {\#15} & {cage14} & \begin{minipage}[s]{0.08\columnwidth}\raisebox{-.5\height}{\includegraphics[width=\linewidth]{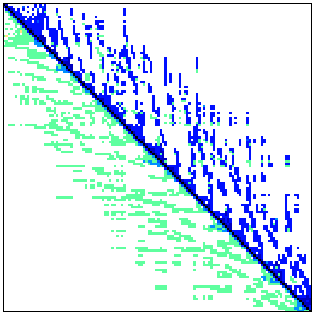}}\end{minipage} & {\makecell{1.5M\\27.1M}} & {\makecell{18.0\\5.37}} & {\#16} & {\makecell{webba\\se-1M}} & \begin{minipage}[s]{0.08\columnwidth}\raisebox{-.5\height}{\includegraphics[width=\linewidth]{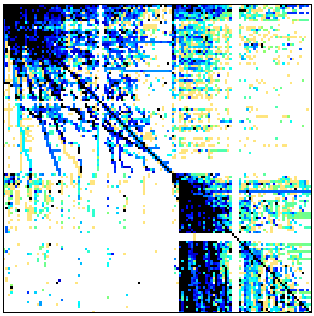}}\end{minipage} & {\makecell{1M\\3.1M}} & {\makecell{3.1\\25.35}}  \\
       \hline
       
\end{tabular}
\vspace{-1em}
\end{table*}

{\bf Memory complexity comparison.} The memory complexity can be divided into two parts: the storage of input matrices and intermediate results. The output matrix, stored in the off-chip memory with COO format, is not considered in the memory complexity analysis. Both COO-SPLIM and SPLIM require $2NK$ memory space to store the compression formats of two input matrices, resulting in an $\mathbf{O}(N)$ memory complexity. During calculations, COO-SPLIM decompresses COO formats to dense matrices, performing the vector multiplication between two $N\times N$ dense matrices and generating $N\times N$ intermediate results, i.e., $\mathbf{O}(N^2)$ for intermediate results. SPLIM adopts the ELLPACK-based computation paradigm, enabling direct SpGEMM using the compressed ELLPACK format. In each iteration of ELLPACK-based vector multiplication, $K\times$ PEs generate $N \times K$ intermediate elements, resulting in a total of $N\times K^2$ intermediate elements for $K\times$ iterations. Thus, SPLIM requires $NK^2$ memory space to store the intermediate matrices, resulting in an $\mathbf{O}(NK^2)$ memory complexity. 


{\bf Time complexity comparison.} COO-SPLIM and SPLIM differ in their computation paradigm, with COO-SPLIM following $alignment \to calculation$ and SPLIM following $calculation \to alignment$, as shown in Figure~\ref{unmatch} and Figure~\ref{vm}. Let's first consider the {\em coordinates alignment} phase. In COO-SPLIM, the coordinates alignment requires $NK\times NK$ search operations, where each COO format of two input matrices contains $N\times K$ non-zero values. Hence, the coordinates alignment has a time complexity of $\mathbf{O}(N^2K^2)$. Similarly, in SPLIM, the vector multiplication generates $NK^2$ intermediate results, and it takes $N\times$ in-situ search iterations on these intermediate results to generate all rows of the output matrix, resulting in the time complexity of $\mathbf{O}(N^2K^2)$.

Let's examine the {\em vector multiplication} phase. In COO-SPLIM, one SpMV iteration in Figure~\ref{unmatch} (c) requires $N\times N$ scalar multiplications. With $N\times$ SpMV iterations for SpGEMM, the overall time complexity is $\mathbf{O}(N^3)$. In contrast, SPLIM's vector multiplication in one ELLPACK iteration only involves $N\times K$ scalar multiplications. With $K\times$ ELLPACK iterations to generate all intermediate matrices, the time complexity in SPLIM is $\mathbf{O}(NK^2)$. Comparing to $\mathbf{O}(N^3)$ complexity of COO-SPLIM, SPLIM has a lower complexity of $O(NK^2)$. 




\section{Methodology}
\label{meth}
{\bf Datasets.} We assess SPLIM's performance by conducting $A\times A^\mathsf{T}$ computations on 16 real-world sparse matrices obtained from the SuiteSparse Matrix Collection~\cite{Davis2011}. These matrices represent diverse application domains, including graph processing, scientific computing, and circuit simulation. A comprehensive overview of these matrices is provided in Table~\ref{tab:matrices}, where we list their names, along with corresponding plots, and designate them with unique matrix IDs ranging from $\#1$ to $\#16$. Additionally, we furnish the number of rows ({\bf Dim.}), the count of non-zero values ($nnz$), the average number of $nnz$ per row ($nnz_{av}$), and the standard deviation of $nnz_{av}$ ($\sigma$).

{\bf Baseline and comparison platforms.} We compare SPLIM to the NVIDIA RTX A6000 GPU, equipped with 46GB memory, 300W TDP, and running on CUDA v11.6. For the GPU baseline, we utilize the cuSPARSE~\cite{cusparse2010} library to conduct SpGEMM computations. We also compare SPLIM with state-of-the-art ASIC-based sparse matrix accelerator SAM~\cite{Hsu2023}, conducting the SpGEMM kernel of their open source project. Furthermore, we benchmark SPLIM against two contemporary PIM-based sparse matrix multiplication accelerators: SpaceA~\cite{Xie2021} and ReFlip~\cite{Huang2022}. We extract the SpMV kernel of SpaceA and ReFlip as descripted in their paper. Then, we extend their methodologies to support SpGEMM through multiple SpMV iterations. The results of SpaceA is obtained with the Ramulator-PIM~\cite{ramulator2015}. We contact the authors of ReFlip to obtain their in-house simulator.

\begin{table}[t]
\centering
\tabcolsep=0.12cm
    \caption{SPLIM Configurations}
    \small
    \renewcommand\arraystretch{1}
    \label{tab:splim}
    \tabcolsep=0.05cm
    \begin{tabular}{|C{1.56cm}||C{1.65cm}|c|c|c|}  
    \cline{1-5}
    
       {\bf Component} & {\bf Area (mm$^2$) } & {\bf Power (mW)} & {\bf Params.} & {\bf Spec.}\\
       \hline \hline
       \multicolumn{5}{|c|}{ELL-PE properties} \\ \hline
       \multirow{3}*{\makecell{ReRAM \\ Arrays}} & \multirow{3}*{3.45} & \multirow{3}*{6.14K} & {Bit per Cell} & {1} \\
       \cline{4-5}
       {} & {} & {} & {Size} & {1024$\times$1024} \\
       \cline{4-5}
       {} & {} & {} & {Total} & {1000} \\
       \hline
       {Buffers} & {0.16} & {79.4} & {Size} & {128KB} \\
       \hline
       {Accumu.} & {0.00024} & {0.2} & {Total} & {1} \\
       \hline
       {{\bf PE total}} & {3.62} & {6.22K} & {Size} & {128.2MB} \\
       \hline\hline
        
       \multirow{2}*{\bf PEs} & \multirow{2}*{115.8} & \multirow{2}*{199.1K} & {Total} & {32} \\
       \cline{4-5}
       {} & {} & {} & {Size} & {4.1GB} \\
       \hline
       {Controller} & {0.013} & {207.8} & {Total} & {1} \\
       \hline
       {{\bf SPLIM}} & {115.8} & {199.3K} & {Size} & {4.1GB} \\
       \hline
\end{tabular}
\end{table}

{\bf SPLIM configurations.} SPLIM is configured with 32 {\em Processing Elements} (PEs) to store the column vectors of input matrices. In cases where the number of column vectors exceeds 32, multiple vectors are stored in a single PE. To manage large input matrices and accommodate intermediate results, SPLIM processes input matrices in batches. Each PE consists of 1000 $1024\times 1024$ memristor arrays. Utilizing 32 memristor cells to store one float32 number, a $1024\times 1024$ memristor array can hold 32 32-bit column vectors (32 columns per vector). Cross-array transfers of RowClone use 1000GB/s On-Chip Interconnect (OCI)~\cite{Jouppi2021}.

Table~\ref{tab:splim} presents the details of one memristor array. We utilize 1GHz {\em one transistor and one memristor} (1T1M) ReRAM arrays for digital in-situ computing. The area and power configurations of the 1T1M memristor array are derived from FloatPIM~\cite{Imani2019}. ReRAM arrays enable column-parallel read/write operations. Additionally, the area and power of on-chip buffers and I/O interfaces are obtained using CACTI 6.5~\cite{Muralimanohar09}. The DRVs of memristor array is implemented using 1-bit DAC from~\cite{Saberi11}. To evaluate SPLIM's latency and energy consumption, we modify ZSim~\cite{ZSim2013} based on the mathematical proof for PUM-based cycle-accurate simulation~\cite{sim2015}.

{\bf Declaration of iso-area.} All non-PUM platforms~\cite{Hsu2023, Xie2021} have two parts of area, i.e., (i) on-chip logic area and (ii) DRAM area. PUM solution~\cite{Huang2022} and SPLIM have only one area because computation and storage are both in ReRAM (iii). It is unfair to SPLIM if we keep (i) and (iii) the same. It's also unfair to other comparison platforms if we keep (ii) the same as (iii) because ReRAM has higher memory density. For a fair comparison, we configure equal-capacity for non-PUM platforms and iso-area for PUM platforms. Specifically, we keep the memory capacity the same as SPLIM for SAM~\cite{Hsu2023} and SpaceA~\cite{Xie2021} while using 3 ReFlip chips~\cite{Huang2022} to maintain the same area with SPLIM.


\begin{figure}[t]
\centering
\includegraphics[width=8.5cm]{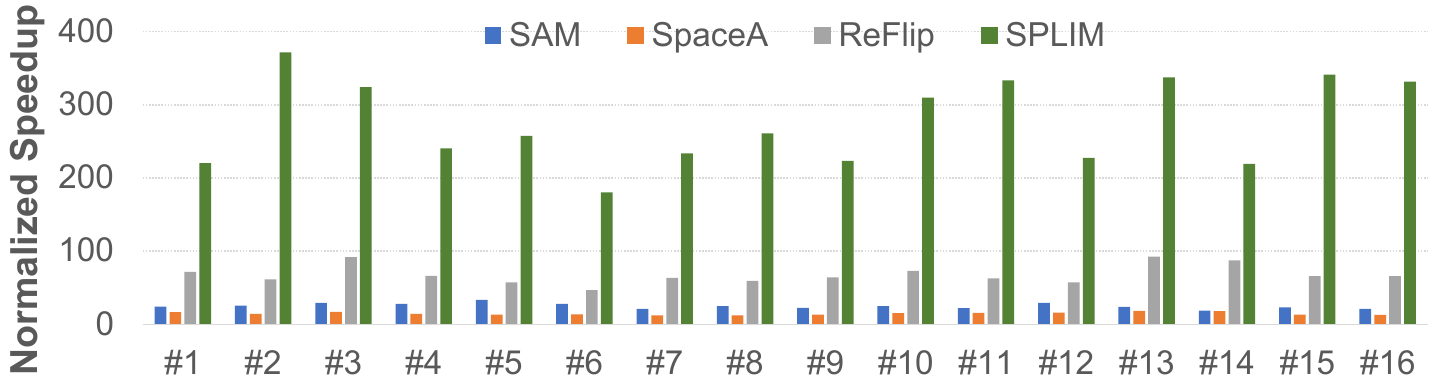}
\caption{Performance comparison between GPU baseline, SAM, SpaceA, ReFlip, and SPLIM (normalized to GPU)}
\label{performance}
\vspace{-1em}
\end{figure}


\section{Evaluation}
\label{exper}

\subsection{Performance and Energy Efficiency}
\begin{figure}[t]
\centering
\includegraphics[width=8.5cm]{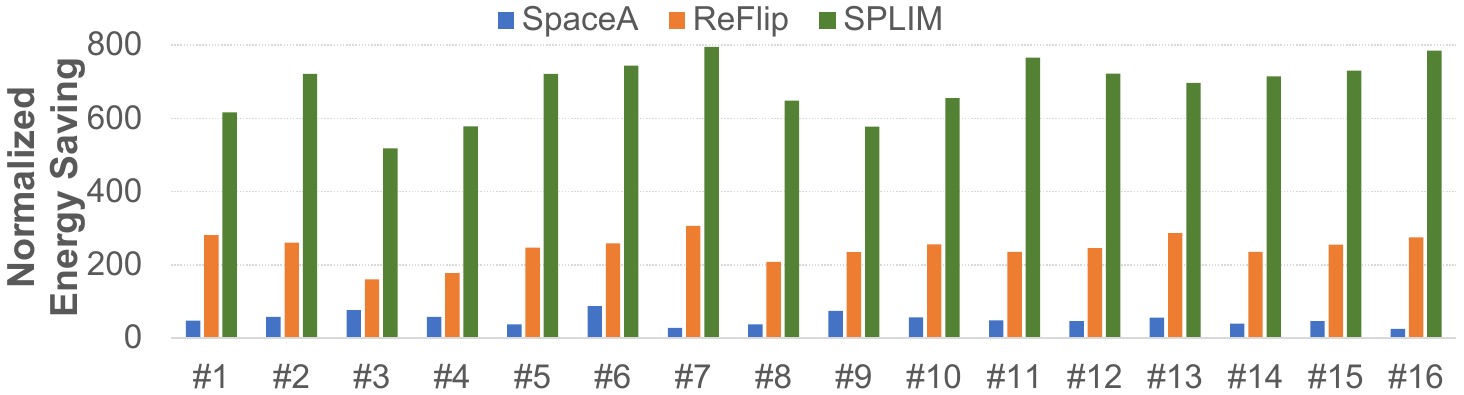}
\caption{Energy saving comparison between GPU baseline, SpaceA, ReFlip, and SPLIM (normalized to GPU)}
\label{energy}
\vspace{-1em}
\end{figure}

We devise experiments to evaluate SPLIM's overall performance and energy efficiency in contrast to GPU baseline, SAM~\cite{Hsu2023}, SpaceA~\cite{Xie2021}, and ReFlip~\cite{Huang2022}. The execution time and energy outcomes are depicted in Figure~\ref{performance} and Figure~\ref{energy}, respectively, with normalization to the GPU. The point-to-point comparison and analysis are detailed below.

{\bf SPLIM vs. GPU.} SPLIM demonstrates a remarkable performance advantage over the GPU baseline, achieving an average performance improvement of 275.74$\times$ and energy savings of 687.19$\times$ across all 16 input matrices. The superiority of SPLIM over the GPU baseline can be attributed to several factors. First, SPLIM effectively mitigates off-chip data movement issues, which are recognized bottlenecks in conventional hardware like GPUs~\cite{Xie2021}. Second, the utilization of the in-situ computing platform exposes SPLIM with million-level row parallelism. Finally, SPLIM adopts SCCP and in-situ search for SpGEMM, eliminating significant random access overhead compared to GPU baseline. The heightened energy efficiency of SPLIM, as compared to the GPU, can be attributed to its avoidance of energy consumption through off-chip transfer and random data access. Additionally, SPLIM's in-situ computing strategy substantially reduces data transfers between the storage and computation units, thus leading to reduced energy consumption. Finally, SPLIM has million-level row parallelism, exposing higher parallelism compared to GPU.

{\bf SPLIM vs. SAM.} SPLIM demonstrates an average performance improvement of 11.08$\times$ when compared to the state-of-the-art ASIC-based SpGEMM accelerator, SAM~\cite{Hsu2023}. As a memory-processor separated architecture, SAM takes many latency to random access non-zero values from the off-chip DRAM. These off-chip data transfers significantly impede SAM performance. Second, SAM still relies on the on-chip scheduler for coordinates alignment. Therefore, scheduling large amount of irregular data makes the scheduler a performance bottleneck for SAM. Conversely, SPLIM's exceptional performance can be attributed to two point-to-point pivotal factors. First, SPLIM operates as a PUM-based accelerator, negating the need for extensive off-chip DRAM access, reducing latency and energy consumption at the same time. Second, SPLIM adopts SCCP and in-situ search rather than on-chip scheduler for SpGEMM, removing the random on-chip scheduling bottleneck with PUM-friendly in-situ operations.

{\bf SPLIM vs. SpaceA.} SPLIM demonstrates an average performance improvement of 19.73$\times$ and achieves energy savings of 13.4$\times$ when compared to the PIM-based SpGEMM accelerator, SpaceA~\cite{Xie2021}. SPLIM outperforms SpaceA in terms of both performance and energy efficiency, benefiting from its PUM-based design that inherently offers enhanced computational parallelism compared to the PIM-based SpaceA. In contrast to SpaceA, which requires scheduling for both input matrices and a significant volume of intermediate results, SPLIM adopts PUM-friendly approaches for intermediate result storage and processing, effectively reducing on-chip memory access overhead. This reduction in memory access overhead enhances both computational and energy efficiency. The innovative utilization of in-situ search operations for coordinate alignment further enhances SPLIM's efficiency, enabling directed read/write operations and mitigating inefficient memory access during intermediate results accumulation.

{\bf SPLIM vs. ReFlip.} SPLIM showcases an average performance improvement of 3.94$\times$ and energy savings of 2.81$\times$ when comparing to the PUM-based accelerator ReFlip~\cite{Huang2022}. Although ReFlip also adopts PUM-based platform, it does not fully utilize the potential of in-situ computing for the following reasons. ReFlip employs a novel data mapping strategy for their SpMV kernel, i.e., storing dense vectors rather than sparse matrices to PUM platform. Their data mapping method performs well for SpMV kernel, because the dense vectors stored in PUM do not introduce any zeros. However, their mapping method fails for processing SpGEMM because there is no dense vectors any more. To this end, ReFlip suffers the same problem as GraphR (Figure~\ref{unmatch}). First, ReFlip needs an additional decompression phase that converts COOs format (for storage) into a dense decompressed format (for computation). This decompression incurs extra data read/write operations between storage and computation formats. Furthermore, ReFlip conducts matrix multiplication in a decompressed dense format, reintroducing zeros and thereby diminishing the array utilization of the PUM platform. In contrast, SPLIM directly employs the compressed ELLPACK format for both storage and computation, eliminating the need for remapping and significantly reducing data read/write operations. The direct use of compressed format for computation mitigates the impact of zeros, increasing array utilization and reduced iterations on the PUM platform. Moreover, SPLIM's adoption of digital in-situ computing obviates the need for intricate peripheral circuits, such as {\em analog-to-digital converters} (ADC) and shift-adders, thereby contributing to improved energy efficiency. Although SPLIM greatly improves hardware utilization, the achieved acceleration is only 3.94$\times$. That is because ReFlip adopts multi-level analog in-situ platform, which has higher density and parallelism than SPLIM's digital in-situ platform.

\subsection{Efficiency of Computation Paradigm}
To assess the effectiveness of the proposed computation paradigm, we establish a sister comparison platform for SPLIM (Section~\ref{complexity}), denoted as COO-SPLIM. While COO-SPLIM maintains identical configurations to SPLIM, the key distinction lies in its computation paradigm, wherein it employs the COO storage format. To mitigate storage-related challenges stemming from decompression, we configure COO-SPLIM to handle the input matrix in batches, thereby processing individual sub-matrices sequentially.


\begin{figure}[t]
\centering
\subfloat[]{
\begin{minipage}[t]{0.489\linewidth}
\centering
\includegraphics[width=1.55in]{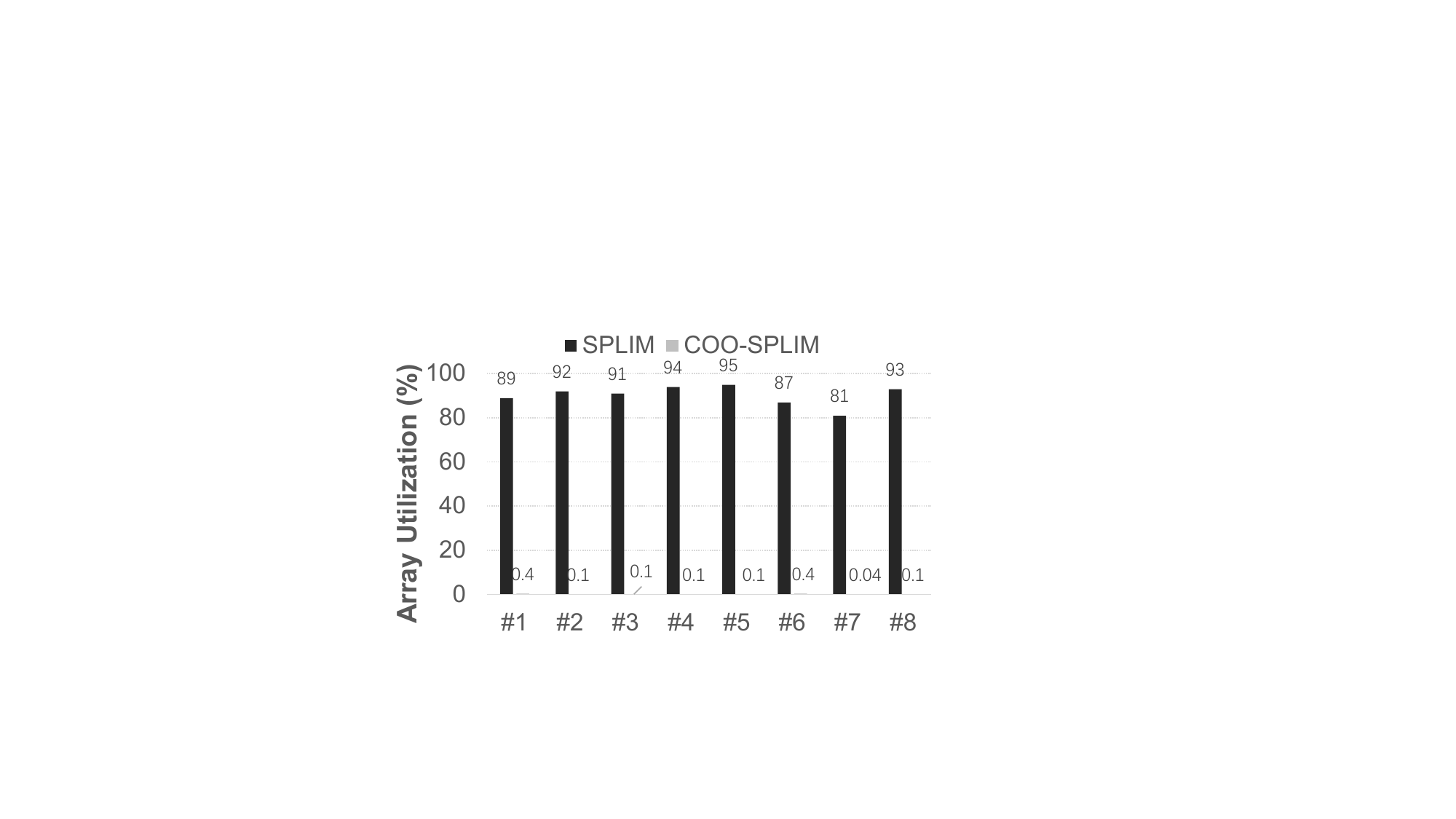}
\end{minipage}
}%
\subfloat[]{
\begin{minipage}[t]{0.489\linewidth}
\centering
\includegraphics[width=1.5in]{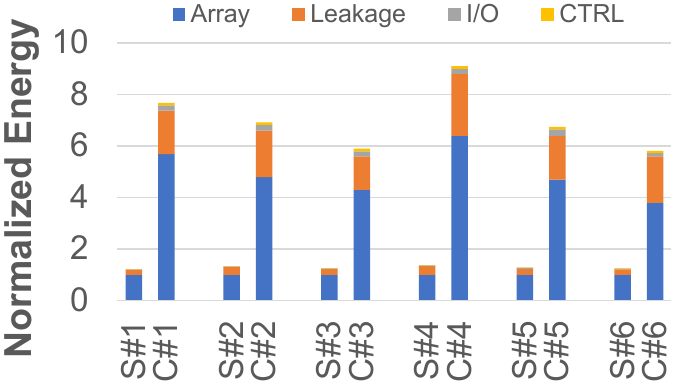}
\end{minipage}
}%
\centering
\caption{(a) Array utilization comparison between SPLIM and COO-SPLIM, and (b) an energy breakdown of SPLIM and COO-SPLIM, with normalization to SPLIM's array energy. In this context, "S\#1" denotes SPLIM processing matrix\#1, and "C\#1" signifies COO-SPLIM processing matrix\#1.}
\label{sister}
\vspace{-1em}
\end{figure}

{\bf Array utilization.} We define array utilization as the ratio of {\em the number of valid ReRAM rows} to {\em the total number of rows}. Figure~\ref{sister} illustrates the array utilization comparison between SPLIM and COO-SPLIM. Notably, SPLIM outperforms COO-SPLIM, exhibiting an average utilization enhancement of 557$\times$. The discernable performance disparity between SPLIM and COO-SPLIM can be attributed to the following reasons. COO-SPLIM adopts an additional decompression phase to convert the COO storage format into a decompressed computation format. This decompression step involves many zeros to the computation region, thereby introducing supplementary invalid computations. Conversely, SPLIM directly employs the compressed ELLPACK format for both storage and calculation, eliminating the need for remapping between different regions. As a result, SPLIM effectively reduces introducing zeros, leading to increase valid computations.

{\bf Energy breakdown.} Figure~\ref{sister} (b) depicts the energy breakdown of both SPLIM and COO-SPLIM. The energy consumption is categorized into four distinct components. The term ``Array" signifies the energy expended by the memristor arrays, while ``Leakage" accounts for the energy dissipation resulting from array current flowing into the {\em ground} (GND), attributed to high resistance cells (`0'). ``I/O" and ``CTRL" correspond to the energy consumption of the I/O interface and the controller, respectively. Notably, COO-SPLIM registers higher energy consumption in ``Array" due to the necessity for a greater number of vector multiplication iterations, consequently requiring more frequent activation of DRVs. Additionally, COO-SPLIM exhibits increased leakage current to the GND owing to the presence of numerous zeros. Both SPLIM and COO-SPLIM demonstrate I/O and CTRL energy consumption of less than 4\%, benefiting from the good energy efficiency of digital in-situ computing.


\subsection{Sensitivity Study}
The sparsity of a sparse matrix and the distribution pattern of its non-zero values will influence the performance of SpGEMM. In this context, we denote $\tau$ as the matrix's sparsity, defined as the ratio of non-zero elements to the total number of elements in the sparse matrix ($\frac{nnz}{Dim^2}$). To gauge the distribution of non-zero elements, we employ the standard deviation ($\sigma$) of the non-zero elements per row. We conduct experiments aimed at examining the impact of variations in $\tau$ and $\sigma$.

\begin{figure}[t]
\centering
\includegraphics[width=8.5cm]{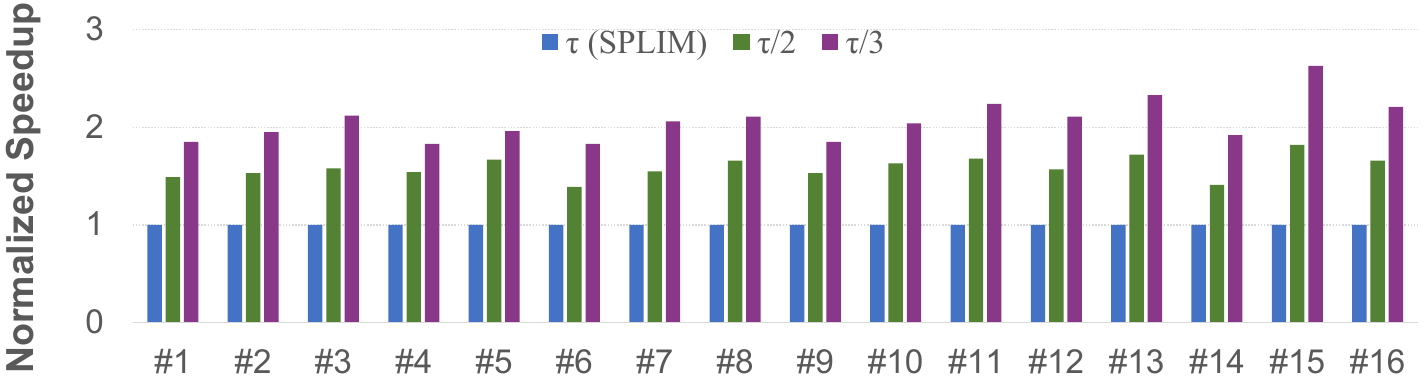}
\caption{Performance comparison of SPLIM processing input matrices with various sparsity (normalized to $\tau$)}
\label{tau}
\vspace{-1em}
\end{figure}

{\bf Impact of matrix sparsity $\tau$.} We configure three levels of input matrix sparsity: $\tau$, $\frac{\tau}{2}$, and $\frac{\tau}{3}$. The sparsity levels of $\frac{\tau}{2}$ and $\frac{\tau}{3}$ are achieved by randomly removing $\frac{1}{2}$ and $\frac{2}{3}$ of the non-zero elements from the sparse matrices, respectively. In Figure~\ref{tau}, we depict the normalized speedup of SPLIM when processing sparse matrices of varying sparsity. SPLIM exhibits improved performance as matrix sparsity increases, all while maintaining the same matrix dimensions. Specifically, transitioning from sparsity $\tau$ to $\frac{\tau}{2}$ results in a 39.6\% reduction in SPLIM's execution time. This trend can be attributed to SPLIM's utilization of the ELLPACK format, effectively condensing all non-zero values. As matrix sparsity diminishes, the number of ELLPACK vectors also decreases, subsequently leading to fewer iterations of vector-vector multiplications.

\begin{figure}[t]
\centering
\includegraphics[width=8.5cm]{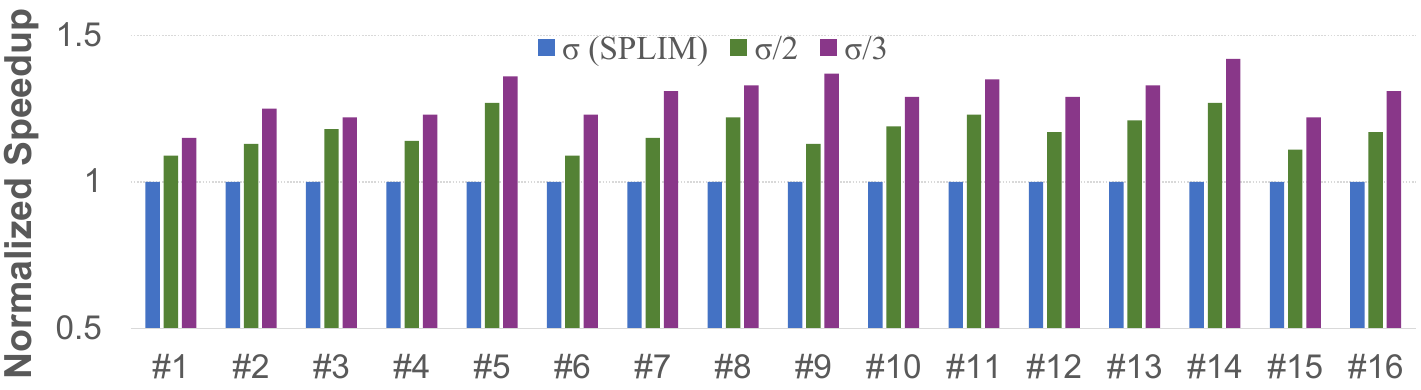}
\caption{Performance comparison of SPLIM processing input matrices with various standard deviation (normalized to $\sigma$)}
\label{sigma}
\vspace{-1em}
\end{figure}

{\bf Impact of standard deviation $\sigma$.} We vary the standard deviation across three levels: $\sigma$, $\frac{\sigma}{2}$, and $\frac{\sigma}{3}$. By redistributing non-zero elements from rows with higher $nnz$ to rows with lower $nnz$, sparse matrices are created with reduced standard deviations of $\frac{\sigma}{2}$ and $\frac{\sigma}{3}$. Figure~\ref{sigma} illustrates the normalized speedup achieved by SPLIM when processing input matrices with varying standard deviations. Notably, SPLIM demonstrates heightened performance as the standard deviation diminishes. A decrease in standard deviation signifies a narrower disparity in non-zero element counts among rows, leading to a reduction in ELLPACK format zeros. The diminished presence of zeros within the ELLPACK format contributes to an enhanced level of array utilization within SPLIM, thus reducing the number of iterations and latency.

\begin{figure}[t]
\centering
\includegraphics[width=8.5cm]{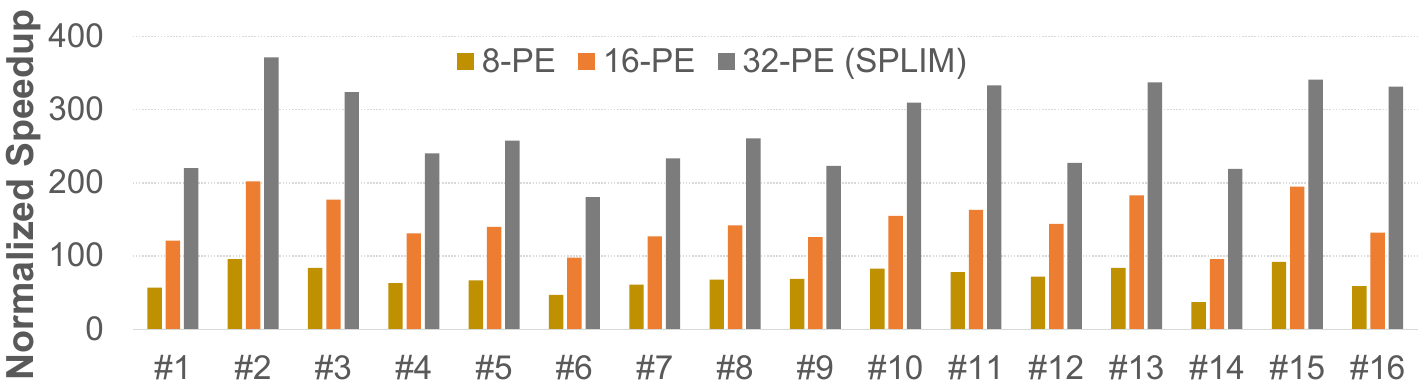}
\caption{Speedups comparison of different PE configurations}
\label{sca-h}
\vspace{-1em}
\end{figure}

\subsection{Scalability}
The scalability of SPLIM is evidenced through incremental augmentation of processing elements (PEs), as depicted in Figure~\ref{sca-h}. When equipped with 32 PEs, SPLIM achieves average speedups of 3.84$\times$ and 1.83$\times$ in comparison to configurations with 8 PEs and 16 PEs, respectively. This performance trend highlights SPLIM's robust scalability, a quality primarily attributed to the utilization of in-situ computing hardware. The in-situ computational memory cells within SPLIM scales proportionally with the number of PEs, leading to a significant enhancement of computational parallelism. The inclusion of additional PEs also facilitates a more granular partitioning of input matrices, resulting in a more uniform distribution of intermediate results. Importantly, SPLIM circumvents the need for cross-PE transfer of intermediate results, rendering it impervious to potential performance hindrances stemming from such transfers.

\section{Related Work}
{\bf Conventional SpGEMM solutions.} Research into accelerating SpGEMM initially took root on the CPU platform~\cite{Gu2020, Patwary2015}, concentrating on optimizing memory bandwidth utilization by exploiting the data locality of SpGEMM calculations. GPUs, with their higher computing parallelism and memory bandwidth, emerged as promising candidates for SpGEMM acceleration~\cite{Lee2020, Niu2022, Parger2020}. The {\em compressed sparse row} (CSR) format has maintained its dominance as the preferred compression scheme on GPUs, giving rise to a wealth of related research~\cite{Dalton2015, gustavson1978two, Liu2014}. Recent innovations have introduced novel matrix compression formats, such as DCSR~\cite{Buluc2008}, BCSR~\cite{BORSTNIK201447}, HNI~\cite{Park2022}, C$^2$SR~\cite{Srivastava2020}, and Bitmap~\cite{Pentecost2019}, tailored to address challenges like irregular memory access and workload imbalance across processing units. In parallel, FPGA~\cite{Haghi2020, Jamro2014, Lin2013} and ASIC~\cite{Hegde2019, Pal2018, Zhang2020} based SpGEMM accelerators have emerged to address inherent latency-bound inefficiencies associated with limited sparse data reuse. However, the performance of these conventional platforms is constrained as the size of sparse matrices expands, with off-chip memory access emerging as a bottleneck~\cite{Xie2021}. Diverging from previous optimizations for SpGEMM across CPU, GPU, FPGA, and ASIC platforms, the emergence of PIM architecture seeks to alleviate the burden of off-chip transmission overhead.

{\bf PIM accelerators.} PIM platforms integrate logic units within memory to offload access-intensive kernels to memory. This in-memory computing paradigm eliminates off-chip transfers, concurrently enhancing the bandwidth of in-memory logic units. PIM platforms have proven adept at enhancing performance for memory-intensive applications, encompassing domains such as neural networks~\cite{Angizi2018, Azarkhish2018, Gao2017, Gu2021, Liu2018}, blockchain~\cite{Wu2019}, and graph processing~\cite{Ahn2015, Dai2019, Nai2017, Zhang2018}. While these solutions excel in addressing load balance, bandwidth utilization, and data reuse, they grapple with issues like C/A bus conflicts and on-chip scheduling overhead stemming from cross-bank transfers. Our focus is on the PUM architecture, which centers on mitigating cross-bank transfers through the innovative concept of in-situ computing. 

{\bf PUM accelerators.} Similar to PIM, PUM harnesses the fundamental principle of relocating computation to memory, circumventing the need for off-chip access. Rather than integrating logic units, PUM leverages the intrinsic computational capabilities of memory cells, exemplified by technologies like {\em Resistive Random Access Memory} (ReRAM)~\cite{reram2010}, {\em Phase Change Memory} (PCM)~\cite{pcm2010}, {\em Spin-Transfer Torque RAM} (STT-RAM)~\cite{stt2011}, and modified DRAM~\cite{dram2019}. By employing memory cells for direct computation, PUM significantly curtails transmission between storage and computation units, concurrently heightening the attainable computational parallelism at the memory array level. This architectural advantage serves as the bedrock for developing various accelerators catering to applications ranging from neural networks~\cite{Cheng2017, Chi16, Imani2019, Ji2019, Shafiee16, Song2017, Wang2018} to graph processing~\cite{Ghasemi2022, Han2018, Song18, Zheng2020}. Among these domains, graph processing bears resemblance to SpGEMM workloads. While prior solutions yield remarkable performance enhancements, they have yet to unlock the full computational parallelism potential of PUM platforms. This is attributed to the inherent trade-off between array-level parallelism and hardware flexibility within PUM platforms. It is within this context that we introduce SPLIM, aimed at bridging the gap between structured PUM hardware and unstructured SpGEMM computations.

\section{Conclusion}
\label{concl}
The objective of this study is to accelerate the commonly employed SpGEMM. PUM-based platforms offer enhanced parallelism and reduced transmission overhead compared to other platforms, making them a promising candidate for accelerating SpGEMM. Nevertheless, attempting to accelerate unstructured SpGEMM using structured PUM platforms results in sub-optimal performance. To bridge this gap, we introduce SPLIM, a novel co-design SpGEMM accelerator that utilizes PUM platforms. First, we develop an innovative computational paradigm with PUM platforms by transforming SpGEMM into structured ELLPACK-based vector multiplication. Second, we introduce a search-based approach for coordinates alignment, converting unstructured accumulation into in-situ search and regular read operations. Finally, we conduct a series of experiments to assess the efficacy of SPLIM. The experimental outcomes demonstrate that SPLIM exhibits remarkable performance and energy efficiency compared to state-of-the-art accelerators.


\bibliographystyle{IEEEtranS}
\bibliography{refs}

\begin{thebibliography}{10}
\providecommand{\url}[1]{#1}
\csname url@samestyle\endcsname
\providecommand{\newblock}{\relax}
\providecommand{\bibinfo}[2]{#2}
\providecommand{\BIBentrySTDinterwordspacing}{\spaceskip=0pt\relax}
\providecommand{\BIBentryALTinterwordstretchfactor}{4}
\providecommand{\BIBentryALTinterwordspacing}{\spaceskip=\fontdimen2\font plus
\BIBentryALTinterwordstretchfactor\fontdimen3\font minus \fontdimen4\font\relax}
\providecommand{\BIBforeignlanguage}[2]{{%
\expandafter\ifx\csname l@#1\endcsname\relax
\typeout{** WARNING: IEEEtranS.bst: No hyphenation pattern has been}%
\typeout{** loaded for the language `#1'. Using the pattern for}%
\typeout{** the default language instead.}%
\else
\language=\csname l@#1\endcsname
\fi
#2}}
\providecommand{\BIBdecl}{\relax}
\BIBdecl

\bibitem{Ahn2015}
\BIBentryALTinterwordspacing
J.~Ahn, S.~Hong, S.~Yoo, O.~Mutlu, and K.~Choi, ``\BIBforeignlanguage{en-US}{A scalable processing-in-memory accelerator for parallel graph processing},'' in \emph{\BIBforeignlanguage{en-US}{Proceedings of the 42nd Annual International Symposium on Computer Architecture}}, Jun 2015. [Online]. Available: \url{http://dx.doi.org/10.1145/2749469.2750386}
\BIBentrySTDinterwordspacing

\bibitem{Akbudak2018}
\BIBentryALTinterwordspacing
K.~Akbudak, O.~Selvitopi, and C.~Aykanat, ``Partitioning models for scaling parallel sparse matrix-matrix multiplication,'' \emph{ACM Trans. Parallel Comput.}, vol.~4, no.~3, jan 2018. [Online]. Available: \url{https://doi.org/10.1145/3155292}
\BIBentrySTDinterwordspacing

\bibitem{reram2010}
H.~Akinaga and H.~Shima, ``Resistive random access memory ({ReRAM}) based on metal oxides,'' \emph{Proceedings of the IEEE}, vol.~98, no.~12, pp. 2237--2251, 2010.

\bibitem{dram2019}
M.~F. Ali, A.~Jaiswal, and K.~Roy, ``In-memory low-cost bit-serial addition using commodity {DRAM} technology,'' \emph{IEEE Transactions on Circuits and Systems I: Regular Papers}, vol.~67, no.~1, pp. 155--165, 2020.

\bibitem{Angizi2018}
\BIBentryALTinterwordspacing
S.~Angizi, Z.~He, A.~S. Rakin, and D.~Fan, ``\BIBforeignlanguage{en-US}{Cmp-pim: an energy-efficient comparator-based processing-in-memory neural network accelerator},'' in \emph{\BIBforeignlanguage{en-US}{Proceedings of the 55th Annual Design Automation Conference}}, Jun 2018. [Online]. Available: \url{http://dx.doi.org/10.1145/3195970.3196009}
\BIBentrySTDinterwordspacing

\bibitem{Azad2016}
\BIBentryALTinterwordspacing
A.~Azad, G.~Ballard, A.~Bulu\c{c}, J.~Demmel, L.~Grigori, O.~Schwartz, S.~Toledo, and S.~Williams, ``Exploiting multiple levels of parallelism in sparse matrix-matrix multiplication,'' \emph{SIAM Journal on Scientific Computing}, vol.~38, no.~6, pp. C624--C651, 2016. [Online]. Available: \url{https://doi.org/10.1137/15M104253X}
\BIBentrySTDinterwordspacing

\bibitem{Azarkhish2018}
\BIBentryALTinterwordspacing
E.~Azarkhish, D.~Rossi, I.~Loi, and L.~Benini, ``\BIBforeignlanguage{en-US}{Neurostream: Scalable and energy efficient deep learning with smart memory cubes},'' \emph{\BIBforeignlanguage{en-US}{IEEE Transactions on Parallel and Distributed Systems}}, vol.~29, no.~2, p. 420–434, Feb 2018. [Online]. Available: \url{http://dx.doi.org/10.1109/tpds.2017.2752706}
\BIBentrySTDinterwordspacing

\bibitem{Hypergraph2016}
\BIBentryALTinterwordspacing
G.~Ballard, A.~Druinsky, N.~Knight, and O.~Schwartz, ``Hypergraph partitioning for sparse matrix-matrix multiplication,'' \emph{ACM Trans. Parallel Comput.}, vol.~3, no.~3, dec 2016. [Online]. Available: \url{https://doi.org/10.1145/3015144}
\BIBentrySTDinterwordspacing

\bibitem{Ballard2016}
\BIBentryALTinterwordspacing
G.~Ballard, C.~Siefert, and J.~Hu, ``Reducing communication costs for sparse matrix multiplication within algebraic multigrid,'' \emph{SIAM Journal on Scientific Computing}, vol.~38, no.~3, pp. C203--C231, 2016. [Online]. Available: \url{https://doi.org/10.1137/15M1028807}
\BIBentrySTDinterwordspacing

\bibitem{Bell2012}
\BIBentryALTinterwordspacing
N.~Bell, S.~Dalton, and L.~N. Olson, ``Exposing fine-grained parallelism in algebraic multigrid methods,'' \emph{SIAM Journal on Scientific Computing}, vol.~34, no.~4, pp. C123--C152, 2012. [Online]. Available: \url{https://doi.org/10.1137/110838844}
\BIBentrySTDinterwordspacing

\bibitem{Bernabeu2015}
S.~R. Bernabeu, V.~Puzyrev, M.~Hanzich, and S.~Fernandez, ``Efficient sparse matrix-vector multiplication for geophysical electromagnetic codes on xeon phi coprocessors,'' in \emph{Second EAGE Workshop on High Performance Computing for Upstream}, vol. 2015, no.~1.\hskip 1em plus 0.5em minus 0.4em\relax European Association of Geoscientists \& Engineers, 2015, pp. 1--5.

\bibitem{Compress2014}
\BIBentryALTinterwordspacing
U.~Borštnik, J.~VandeVondele, V.~Weber, and J.~Hutter, ``Sparse matrix multiplication: The distributed block-compressed sparse row library,'' \emph{Parallel Computing}, vol.~40, no.~5, pp. 47--58, 2014. [Online]. Available: \url{https://www.sciencedirect.com/science/article/pii/S0167819114000428}
\BIBentrySTDinterwordspacing

\bibitem{BORSTNIK201447}
\BIBentryALTinterwordspacing
U.~Borštnik, J.~VandeVondele, V.~Weber, and J.~Hutter, ``Sparse matrix multiplication: The distributed block-compressed sparse row library,'' \emph{Parallel Computing}, vol.~40, no.~5, pp. 47--58, 2014. [Online]. Available: \url{https://www.sciencedirect.com/science/article/pii/S0167819114000428}
\BIBentrySTDinterwordspacing

\bibitem{Buluc2008}
A.~Buluc and J.~R. Gilbert, ``On the representation and multiplication of hypersparse matrices,'' in \emph{2008 IEEE International Symposium on Parallel and Distributed Processing}, 2008, pp. 1--11.

\bibitem{Gilbert2011}
\BIBentryALTinterwordspacing
A.~Buluç and J.~R. Gilbert, ``The combinatorial blas: design, implementation, and applications,'' \emph{The International Journal of High Performance Computing Applications}, vol.~25, no.~4, pp. 496--509, 2011. [Online]. Available: \url{https://doi.org/10.1177/1094342011403516}
\BIBentrySTDinterwordspacing

\bibitem{Challapalle20}
N.~Challapalle, S.~Rampalli, L.~Song, N.~Chandramoorthy, K.~Swaminathan, J.~Sampson, Y.~Chen, and V.~Narayanan, ``Gaas-x: Graph analytics accelerator supporting sparse data representation using crossbar architectures,'' in \emph{2020 ACM/IEEE 47th Annual International Symposium on Computer Architecture (ISCA)}.\hskip 1em plus 0.5em minus 0.4em\relax IEEE, 2020, pp. 433--445.

\bibitem{Cheng2017}
\BIBentryALTinterwordspacing
M.~Cheng, L.~Xia, Z.~Zhu, Y.~Cai, Y.~Xie, Y.~Wang, and H.~Yang, ``\BIBforeignlanguage{en-US}{Time: A training-in-memory architecture for memristor-based deep neural networks},'' in \emph{\BIBforeignlanguage{en-US}{Proceedings of the 54th Annual Design Automation Conference 2017}}, Jun 2017. [Online]. Available: \url{http://dx.doi.org/10.1145/3061639.3062326}
\BIBentrySTDinterwordspacing

\bibitem{Chi16}
\BIBentryALTinterwordspacing
P.~Chi, S.~Li, C.~Xu, T.~Zhang, J.~Zhao, Y.~Liu, Y.~Wang, and Y.~Xie, ``Prime: A novel processing-in-memory architecture for neural network computation in reram-based main memory,'' in \emph{Proceedings of the 43rd International Symposium on Computer Architecture}, ser. ISCA '16.\hskip 1em plus 0.5em minus 0.4em\relax IEEE Press, 2016, p. 27–39. [Online]. Available: \url{https://doi.org/10.1109/ISCA.2016.13}
\BIBentrySTDinterwordspacing

\bibitem{Dai2019}
\BIBentryALTinterwordspacing
G.~Dai, T.~Huang, Y.~Chi, J.~Zhao, G.~Sun, Y.~Liu, Y.~Wang, Y.~Xie, and H.~Yang, ``\BIBforeignlanguage{en-US}{Graphh: A processing-in-memory architecture for large-scale graph processing},'' \emph{\BIBforeignlanguage{en-US}{IEEE Transactions on Computer-Aided Design of Integrated Circuits and Systems}}, p. 640–653, Apr 2019. [Online]. Available: \url{http://dx.doi.org/10.1109/tcad.2018.2821565}
\BIBentrySTDinterwordspacing

\bibitem{Dalton2015}
\BIBentryALTinterwordspacing
S.~Dalton, L.~Olson, and N.~Bell, ``Optimizing sparse matrix—matrix multiplication for the gpu,'' \emph{ACM Trans. Math. Softw.}, vol.~41, no.~4, oct 2015. [Online]. Available: \url{https://doi.org/10.1145/2699470}
\BIBentrySTDinterwordspacing

\bibitem{Davis2011}
\BIBentryALTinterwordspacing
T.~A. Davis and Y.~Hu, ``The university of florida sparse matrix collection,'' \emph{ACM Trans. Math. Softw.}, vol.~38, no.~1, dec 2011. [Online]. Available: \url{https://doi.org/10.1145/2049662.2049663}
\BIBentrySTDinterwordspacing

\bibitem{Demirci2020}
G.~V. Demirci and C.~Aykanat, ``Cartesian partitioning models for 2d and 3d parallel spgemm algorithms,'' \emph{IEEE Transactions on Parallel and Distributed Systems}, vol.~31, no.~12, pp. 2763--2775, 2020.

\bibitem{Feng2022}
\BIBentryALTinterwordspacing
S.~Feng, X.~He, K.-Y. Chen, L.~Ke, X.~Zhang, D.~Blaauw, T.~Mudge, and R.~Dreslinski, ``Menda: A near-memory multi-way merge solution for sparse transposition and dataflows,'' in \emph{Proceedings of the 49th Annual International Symposium on Computer Architecture}, ser. ISCA '22.\hskip 1em plus 0.5em minus 0.4em\relax New York, NY, USA: Association for Computing Machinery, 2022, p. 245–258. [Online]. Available: \url{https://doi.org/10.1145/3470496.3527432}
\BIBentrySTDinterwordspacing

\bibitem{Gao2017}
\BIBentryALTinterwordspacing
M.~Gao, J.~Pu, X.~Yang, M.~Horowitz, and C.~Kozyrakis, ``\BIBforeignlanguage{en-US}{Tetris: Scalable and efficient neural network acceleration with 3d memory},'' in \emph{\BIBforeignlanguage{en-US}{Proceedings of the Twenty-Second International Conference on Architectural Support for Programming Languages and Operating Systems}}, Apr 2017. [Online]. Available: \url{http://dx.doi.org/10.1145/3037697.3037702}
\BIBentrySTDinterwordspacing

\bibitem{Ghasemi2022}
S.~A. Ghasemi, B.~Jahannia, and H.~Farbeh, ``Grapha: An efficient reram-based architecture to accelerate large scale graph processing,'' \emph{Journal of Systems Architecture}, vol. 133, p. 102755, 2022.

\bibitem{Giannoula2022}
\BIBentryALTinterwordspacing
C.~Giannoula, I.~Fernandez, J.~G. Luna, N.~Koziris, G.~Goumas, and O.~Mutlu, ``Sparsep: Towards efficient sparse matrix vector multiplication on real processing-in-memory architectures,'' \emph{Proc. ACM Meas. Anal. Comput. Syst.}, vol.~6, no.~1, feb 2022. [Online]. Available: \url{https://doi.org/10.1145/3508041}
\BIBentrySTDinterwordspacing

\bibitem{Gilbert1992}
\BIBentryALTinterwordspacing
J.~R. Gilbert, C.~Moler, and R.~Schreiber, ``Sparse matrices in matlab: Design and implementation,'' \emph{SIAM Journal on Matrix Analysis and Applications}, vol.~13, no.~1, pp. 333--356, 1992. [Online]. Available: \url{https://doi.org/10.1137/0613024}
\BIBentrySTDinterwordspacing

\bibitem{Gondimalla2019}
\BIBentryALTinterwordspacing
A.~Gondimalla, N.~Chesnut, M.~Thottethodi, and T.~N. Vijaykumar, ``Sparten: A sparse tensor accelerator for convolutional neural networks,'' in \emph{Proceedings of the 52nd Annual IEEE/ACM International Symposium on Microarchitecture}.\hskip 1em plus 0.5em minus 0.4em\relax New York, NY, USA: Association for Computing Machinery, 2019, p. 151–165. [Online]. Available: \url{https://doi.org/10.1145/3352460.3358291}
\BIBentrySTDinterwordspacing

\bibitem{ellpack1979}
R.~G. Grimes, D.~R. Kincaid, and D.~M. Young, \emph{ITPACK 2.0 user's guide}.\hskip 1em plus 0.5em minus 0.4em\relax Center for Numerical Analysis, Univ., 1979.

\bibitem{Gu2021}
\BIBentryALTinterwordspacing
P.~Gu, X.~Xie, S.~Li, D.~Niu, H.~Zheng, K.~T. Malladi, and Y.~Xie, ``\BIBforeignlanguage{en-US}{Dlux: A lut-based near-bank accelerator for data center deep learning training workloads},'' \emph{\BIBforeignlanguage{en-US}{IEEE Transactions on Computer-Aided Design of Integrated Circuits and Systems}}, vol.~40, no.~8, p. 1586–1599, Aug 2021. [Online]. Available: \url{http://dx.doi.org/10.1109/tcad.2020.3021336}
\BIBentrySTDinterwordspacing

\bibitem{Gu2020}
\BIBentryALTinterwordspacing
Z.~Gu, J.~Moreira, D.~Edelsohn, and A.~Azad, ``Bandwidth optimized parallel algorithms for sparse matrix-matrix multiplication using propagation blocking,'' in \emph{Proceedings of the 32nd ACM Symposium on Parallelism in Algorithms and Architectures}, ser. SPAA '20.\hskip 1em plus 0.5em minus 0.4em\relax New York, NY, USA: Association for Computing Machinery, 2020, p. 293–303. [Online]. Available: \url{https://doi.org/10.1145/3350755.3400216}
\BIBentrySTDinterwordspacing

\bibitem{Gupta18}
S.~Gupta, M.~Imani, and T.~Rosing, ``Felix: Fast and energy-efficient logic in memory,'' in \emph{2018 IEEE/ACM International Conference on Computer-Aided Design (ICCAD)}, 2018, pp. 1--7.

\bibitem{gustavson1978two}
F.~G. Gustavson, ``Two fast algorithms for sparse matrices: Multiplication and permuted transposition,'' \emph{ACM Transactions on Mathematical Software (TOMS)}, vol.~4, no.~3, pp. 250--269, 1978.

\bibitem{Haghi2020}
P.~Haghi, T.~Geng, A.~Guo, T.~Wang, and M.~Herbordt, ``Fp-amg: Fpga-based acceleration framework for algebraic multigrid solvers,'' in \emph{2020 IEEE 28th Annual International Symposium on Field-Programmable Custom Computing Machines (FCCM)}, 2020, pp. 148--156.

\bibitem{Han2018}
\BIBentryALTinterwordspacing
L.~Han, Z.~Shen, D.~Liu, Z.~Shao, H.~H. Huang, and T.~Li, ``A novel reram-based processing-in-memory architecture for graph traversal,'' \emph{ACM Trans. Storage}, vol.~14, no.~1, feb 2018. [Online]. Available: \url{https://doi.org/10.1145/3177916}
\BIBentrySTDinterwordspacing

\bibitem{Hegde2019}
\BIBentryALTinterwordspacing
K.~Hegde, H.~Asghari-Moghaddam, M.~Pellauer, N.~Crago, A.~Jaleel, E.~Solomonik, J.~Emer, and C.~W. Fletcher, ``Extensor: An accelerator for sparse tensor algebra,'' in \emph{Proceedings of the 52nd Annual IEEE/ACM International Symposium on Microarchitecture}, ser. MICRO '52.\hskip 1em plus 0.5em minus 0.4em\relax New York, NY, USA: Association for Computing Machinery, 2019, p. 319–333. [Online]. Available: \url{https://doi.org/10.1145/3352460.3358275}
\BIBentrySTDinterwordspacing

\bibitem{Hsu2023}
\BIBentryALTinterwordspacing
O.~Hsu, M.~Strange, R.~Sharma, J.~Won, K.~Olukotun, J.~S. Emer, M.~A. Horowitz, and F.~Kj\o{}lstad, ``The sparse abstract machine,'' in \emph{Proceedings of the 28th ACM International Conference on Architectural Support for Programming Languages and Operating Systems, Volume 3}, ser. ASPLOS 2023.\hskip 1em plus 0.5em minus 0.4em\relax New York, NY, USA: Association for Computing Machinery, 2023, p. 710–726. [Online]. Available: \url{https://doi.org/10.1145/3582016.3582051}
\BIBentrySTDinterwordspacing

\bibitem{Huang2022}
Y.~Huang, L.~Zheng, P.~Yao, Q.~Wang, X.~Liao, H.~Jin, and J.~Xue, ``Accelerating graph convolutional networks using crossbar-based processing-in-memory architectures,'' in \emph{2022 IEEE International Symposium on High-Performance Computer Architecture (HPCA)}, 2022, pp. 1029--1042.

\bibitem{Imani2019}
\BIBentryALTinterwordspacing
M.~Imani, S.~Gupta, Y.~Kim, and T.~Rosing, ``Floatpim: In-memory acceleration of deep neural network training with high precision,'' in \emph{Proceedings of the 46th International Symposium on Computer Architecture}, ser. ISCA '19.\hskip 1em plus 0.5em minus 0.4em\relax New York, NY, USA: Association for Computing Machinery, 2019, p. 802–815. [Online]. Available: \url{https://doi.org/10.1145/3307650.3322237}
\BIBentrySTDinterwordspacing

\bibitem{Jamro2014}
E.~Jamro, T.~Pabi{\'s}, P.~Russek, and K.~Wiatr, ``The algorithms for fpga implementation of sparse matrices multiplication,'' \emph{Computing and Informatics}, vol.~33, no.~3, pp. 667--684, 2014.

\bibitem{Jang18}
\BIBentryALTinterwordspacing
B.~C. Jang, Y.~Nam, B.~J. Koo, J.~Choi, S.~G. Im, S.-H.~K. Park, and S.-Y. Choi, ``Memristive logic-in-memory integrated circuits for energy-efficient flexible electronics,'' \emph{Advanced Functional Materials}, vol.~28, no.~2, p. 1704725, 2018. [Online]. Available: \url{https://onlinelibrary.wiley.com/doi/abs/10.1002/adfm.201704725}
\BIBentrySTDinterwordspacing

\bibitem{Ji2019}
\BIBentryALTinterwordspacing
Y.~Ji, Y.~Zhang, X.~Xie, S.~Li, P.~Wang, X.~Hu, Y.~Zhang, and Y.~Xie, ``\BIBforeignlanguage{en-US}{Fpsa: A full system stack solution for reconfigurable reram-based nn accelerator architecture},'' in \emph{\BIBforeignlanguage{en-US}{Proceedings of the Twenty-Fourth International Conference on Architectural Support for Programming Languages and Operating Systems}}, Apr 2019. [Online]. Available: \url{http://dx.doi.org/10.1145/3297858.3304048}
\BIBentrySTDinterwordspacing

\bibitem{Jouppi2021}
N.~P. Jouppi, D.~Hyun~Yoon, M.~Ashcraft, M.~Gottscho, T.~B. Jablin, G.~Kurian, J.~Laudon, S.~Li, P.~Ma, X.~Ma, T.~Norrie, N.~Patil, S.~Prasad, C.~Young, Z.~Zhou, and D.~Patterson, ``Ten lessons from three generations shaped google’s tpuv4i: Industrial product,'' in \emph{2021 ACM/IEEE 48th Annual International Symposium on Computer Architecture (ISCA)}.\hskip 1em plus 0.5em minus 0.4em\relax IEEE, 2021, pp. 1--14.

\bibitem{ramulator2015}
Y.~Kim, W.~Yang, and O.~Mutlu, ``Ramulator: A fast and extensible {DRAM} simulator,'' \emph{IEEE Computer Architecture Letters}, vol.~15, no.~1, pp. 45--49, 2016.

\bibitem{Lee2020}
J.~Lee, S.~Kang, Y.~Yu, Y.-Y. Jo, S.-W. Kim, and Y.~Park, ``Optimization of gpu-based sparse matrix multiplication for large sparse networks,'' in \emph{2020 IEEE 36th International Conference on Data Engineering (ICDE)}, 2020, pp. 925--936.

\bibitem{resqm2020}
H.~Li, H.~Jin, L.~Zheng, and X.~Liao, ``{ReSQM}: Accelerating database operations using {ReRAM}-based content addressable memory,'' \emph{IEEE Transactions on Computer-Aided Design of Integrated Circuits and Systems}, vol.~39, no.~11, pp. 4030--4041, 2020.

\bibitem{Pinatubo2016}
\BIBentryALTinterwordspacing
S.~Li, C.~Xu, Q.~Zou, J.~Zhao, Y.~Lu, and Y.~Xie, ``Pinatubo: A processing-in-memory architecture for bulk bitwise operations in emerging non-volatile memories,'' in \emph{Proceedings of the 53rd Annual Design Automation Conference}, ser. DAC '16.\hskip 1em plus 0.5em minus 0.4em\relax New York, NY, USA: Association for Computing Machinery, 2016. [Online]. Available: \url{https://doi.org/10.1145/2897937.2898064}
\BIBentrySTDinterwordspacing

\bibitem{Lin2013}
C.~Y. Lin, N.~Wong, and H.~K.-H. So, ``Design space exploration for sparse matrix-matrix multiplication on fpgas,'' \emph{International Journal of Circuit Theory and Applications}, vol.~41, no.~2, pp. 205--219, 2013.

\bibitem{Liu2015}
B.~Liu, M.~Wang, H.~Foroosh, M.~Tappen, and M.~Pensky, ``Sparse convolutional neural networks,'' in \emph{Proceedings of the IEEE Conference on Computer Vision and Pattern Recognition (CVPR)}, June 2015.

\bibitem{Liu2018}
\BIBentryALTinterwordspacing
J.~Liu, H.~Zhao, M.~A. Ogleari, D.~Li, and J.~Zhao, ``\BIBforeignlanguage{en-US}{Processing-in-memory for energy-efficient neural network training: a heterogeneous approach},'' in \emph{\BIBforeignlanguage{en-US}{2018 51st Annual IEEE/ACM International Symposium on Microarchitecture (MICRO)}}, Oct 2018. [Online]. Available: \url{http://dx.doi.org/10.1109/micro.2018.00059}
\BIBentrySTDinterwordspacing

\bibitem{Liu2014}
W.~Liu and B.~Vinter, ``An efficient gpu general sparse matrix-matrix multiplication for irregular data,'' in \emph{2014 IEEE 28th International Parallel and Distributed Processing Symposium}, 2014, pp. 370--381.

\bibitem{Lu21}
\BIBentryALTinterwordspacing
L.~Lu, Y.~Jin, H.~Bi, Z.~Luo, P.~Li, T.~Wang, and Y.~Liang, ``Sanger: A co-design framework for enabling sparse attention using reconfigurable architecture,'' in \emph{MICRO-54: 54th Annual IEEE/ACM International Symposium on Microarchitecture}, ser. MICRO '21.\hskip 1em plus 0.5em minus 0.4em\relax New York, NY, USA: Association for Computing Machinery, 2021, p. 977–991. [Online]. Available: \url{https://doi.org/10.1145/3466752.3480125}
\BIBentrySTDinterwordspacing

\bibitem{Lyu2023}
B.~Lyu, M.~Hamdi, Y.~Yang, Y.~Cao, Z.~Yan, K.~Li, S.~Wen, and T.~Huang, ``Efficient spectral graph convolutional network deployment on memristive crossbars,'' \emph{IEEE Transactions on Emerging Topics in Computational Intelligence}, vol.~7, no.~2, pp. 415--425, 2023.

\bibitem{Muralimanohar09}
N.~Muralimanohar, R.~Balasubramonian, and N.~P. Jouppi, ``Cacti 6.0: A tool to model large caches,'' \emph{HP laboratories}, vol.~27, p.~28, 2009.

\bibitem{Nai2017}
\BIBentryALTinterwordspacing
L.~Nai, R.~Hadidi, J.~Sim, H.~Kim, P.~Kumar, and H.~Kim, ``\BIBforeignlanguage{en-US}{Graphpim: Enabling instruction-level pim offloading in graph computing frameworks},'' in \emph{\BIBforeignlanguage{en-US}{2017 IEEE International Symposium on High Performance Computer Architecture (HPCA)}}, Feb 2017. [Online]. Available: \url{http://dx.doi.org/10.1109/hpca.2017.54}
\BIBentrySTDinterwordspacing

\bibitem{cusparse2010}
M.~Naumov, L.~Chien, P.~Vandermersch, and U.~Kapasi, ``Cusparse library,'' in \emph{GPU Technology Conference}, 2010.

\bibitem{Niu2022}
\BIBentryALTinterwordspacing
Y.~Niu, Z.~Lu, H.~Ji, S.~Song, Z.~Jin, and W.~Liu, ``Tilespgemm: A tiled algorithm for parallel sparse general matrix-matrix multiplication on gpus,'' in \emph{Proceedings of the 27th ACM SIGPLAN Symposium on Principles and Practice of Parallel Programming}, ser. PPoPP '22.\hskip 1em plus 0.5em minus 0.4em\relax New York, NY, USA: Association for Computing Machinery, 2022, p. 90–106. [Online]. Available: \url{https://doi.org/10.1145/3503221.3508431}
\BIBentrySTDinterwordspacing

\bibitem{Pal2018}
S.~Pal, J.~Beaumont, D.-H. Park, A.~Amarnath, S.~Feng, C.~Chakrabarti, H.-S. Kim, D.~Blaauw, T.~Mudge, and R.~Dreslinski, ``Outerspace: An outer product based sparse matrix multiplication accelerator,'' in \emph{2018 IEEE International Symposium on High Performance Computer Architecture (HPCA)}, 2018, pp. 724--736.

\bibitem{Parger2020}
\BIBentryALTinterwordspacing
M.~Parger, M.~Winter, D.~Mlakar, and M.~Steinberger, ``Speck: Accelerating gpu sparse matrix-matrix multiplication through lightweight analysis,'' in \emph{Proceedings of the 25th ACM SIGPLAN Symposium on Principles and Practice of Parallel Programming}, ser. PPoPP '20.\hskip 1em plus 0.5em minus 0.4em\relax New York, NY, USA: Association for Computing Machinery, 2020, p. 362–375. [Online]. Available: \url{https://doi.org/10.1145/3332466.3374521}
\BIBentrySTDinterwordspacing

\bibitem{Park2022}
S.~Park, J.-J. Kim, and J.~Kung, ``Autorelax: Hw-sw co-optimization for efficient spgemm operations with automated relaxation in deep learning,'' \emph{IEEE Transactions on Emerging Topics in Computing}, vol.~10, no.~3, pp. 1428--1442, 2022.

\bibitem{Patwary2015}
M.~M.~A. Patwary, N.~R. Satish, N.~Sundaram, J.~Park, M.~J. Anderson, S.~G. Vadlamudi, D.~Das, S.~G. Pudov, V.~O. Pirogov, and P.~Dubey, ``Parallel efficient sparse matrix-matrix multiplication on multicore platforms,'' in \emph{High Performance Computing}, J.~M. Kunkel and T.~Ludwig, Eds.\hskip 1em plus 0.5em minus 0.4em\relax Cham: Springer International Publishing, 2015, pp. 48--57.

\bibitem{Pentecost2019}
\BIBentryALTinterwordspacing
L.~Pentecost, M.~Donato, B.~Reagen, U.~Gupta, S.~Ma, G.-Y. Wei, and D.~Brooks, ``Maxnvm: Maximizing dnn storage density and inference efficiency with sparse encoding and error mitigation,'' in \emph{Proceedings of the 52nd Annual IEEE/ACM International Symposium on Microarchitecture}, ser. MICRO '52.\hskip 1em plus 0.5em minus 0.4em\relax New York, NY, USA: Association for Computing Machinery, 2019, p. 769–781. [Online]. Available: \url{https://doi.org/10.1145/3352460.3358258}
\BIBentrySTDinterwordspacing

\bibitem{Qiu2022}
S.~Qiu, L.~You, and Z.~Wang, ``Optimizing sparse matrix multiplications for graph neural networks,'' in \emph{Languages and Compilers for Parallel Computing}, X.~Li and S.~Chandrasekaran, Eds.\hskip 1em plus 0.5em minus 0.4em\relax Cham: Springer International Publishing, 2022, pp. 101--117.

\bibitem{Saberi11}
M.~Saberi, R.~Lotfi, K.~Mafinezhad, and W.~A. Serdijn, ``Analysis of power consumption and linearity in capacitive digital-to-analog converters used in successive approximation adcs,'' \emph{IEEE Transactions on Circuits and Systems I: Regular Papers}, vol.~58, no.~8, pp. 1736--1748, 2011.

\bibitem{ZSim2013}
\BIBentryALTinterwordspacing
D.~Sanchez and C.~Kozyrakis, ``Zsim: Fast and accurate microarchitectural simulation of thousand-core systems,'' in \emph{Proceedings of the 40th Annual International Symposium on Computer Architecture}, ser. ISCA '13.\hskip 1em plus 0.5em minus 0.4em\relax New York, NY, USA: Association for Computing Machinery, 2013, p. 475–486. [Online]. Available: \url{https://doi.org/10.1145/2485922.2485963}
\BIBentrySTDinterwordspacing

\bibitem{rowclone2013}
\BIBentryALTinterwordspacing
V.~Seshadri, Y.~Kim, C.~Fallin, D.~Lee, R.~Ausavarungnirun, G.~Pekhimenko, Y.~Luo, O.~Mutlu, P.~B. Gibbons, M.~A. Kozuch, and T.~C. Mowry, ``Rowclone: Fast and energy-efficient in-dram bulk data copy and initialization,'' in \emph{Proceedings of the 46th Annual IEEE/ACM International Symposium on Microarchitecture}, ser. MICRO-46.\hskip 1em plus 0.5em minus 0.4em\relax New York, NY, USA: Association for Computing Machinery, 2013, p. 185–197. [Online]. Available: \url{https://doi.org/10.1145/2540708.2540725}
\BIBentrySTDinterwordspacing

\bibitem{Shafiee16}
\BIBentryALTinterwordspacing
A.~Shafiee, A.~Nag, N.~Muralimanohar, R.~Balasubramonian, J.~P. Strachan, M.~Hu, R.~S. Williams, and V.~Srikumar, ``Isaac: A convolutional neural network accelerator with in-situ analog arithmetic in crossbars,'' in \emph{Proceedings of the 43rd International Symposium on Computer Architecture}, ser. ISCA '16.\hskip 1em plus 0.5em minus 0.4em\relax IEEE Press, 2016, p. 14–26. [Online]. Available: \url{https://doi.org/10.1109/ISCA.2016.12}
\BIBentrySTDinterwordspacing

\bibitem{Siemon15}
A.~Siemon, S.~Menzel, R.~Waser, and E.~Linn, ``A complementary resistive switch-based crossbar array adder,'' \emph{IEEE Journal on Emerging and Selected Topics in Circuits and Systems}, vol.~5, no.~1, pp. 64--74, 2015.

\bibitem{stt2011}
C.~W. Smullen, V.~Mohan, A.~Nigam, S.~Gurumurthi, and M.~R. Stan, ``Relaxing non-volatility for fast and energy-efficient {STT-RAM} caches,'' in \emph{Proceedings of 2011 IEEE 17th International Symposium on High Performance Computer Architecture}, 2011, pp. 50--61.

\bibitem{Song2017}
\BIBentryALTinterwordspacing
L.~Song, X.~Qian, H.~Li, and Y.~Chen, ``\BIBforeignlanguage{en-US}{Pipelayer: A pipelined reram-based accelerator for deep learning},'' in \emph{\BIBforeignlanguage{en-US}{2017 IEEE International Symposium on High Performance Computer Architecture (HPCA)}}, Feb 2017. [Online]. Available: \url{http://dx.doi.org/10.1109/hpca.2017.55}
\BIBentrySTDinterwordspacing

\bibitem{Song18}
L.~Song, Y.~Zhuo, X.~Qian, H.~Li, and Y.~Chen, ``Graphr: Accelerating graph processing using reram,'' in \emph{2018 IEEE International Symposium on High Performance Computer Architecture (HPCA)}.\hskip 1em plus 0.5em minus 0.4em\relax IEEE, 2018, pp. 531--543.

\bibitem{Srivastava2020}
N.~Srivastava, H.~Jin, J.~Liu, D.~Albonesi, and Z.~Zhang, ``Matraptor: A sparse-sparse matrix multiplication accelerator based on row-wise product,'' in \emph{2020 53rd Annual IEEE/ACM International Symposium on Microarchitecture (MICRO)}, 2020, pp. 766--780.

\bibitem{Then2014}
\BIBentryALTinterwordspacing
M.~Then, M.~Kaufmann, F.~Chirigati, T.-A. Hoang-Vu, K.~Pham, A.~Kemper, T.~Neumann, and H.~T. Vo, ``The more the merrier: Efficient multi-source graph traversal,'' \emph{Proc. VLDB Endow.}, vol.~8, no.~4, p. 449–460, dec 2014. [Online]. Available: \url{https://doi.org/10.14778/2735496.2735507}
\BIBentrySTDinterwordspacing

\bibitem{Wang2018}
\BIBentryALTinterwordspacing
P.~Wang, Y.~Ji, C.~Hong, Y.~Lyu, D.~Wang, and Y.~Xie, ``Snrram: An efficient sparse neural network computation architecture based on resistive random-access memory,'' in \emph{Proceedings of the 55th Annual Design Automation Conference}, ser. DAC '18.\hskip 1em plus 0.5em minus 0.4em\relax New York, NY, USA: Association for Computing Machinery, 2018. [Online]. Available: \url{https://doi.org/10.1145/3195970.3196116}
\BIBentrySTDinterwordspacing

\bibitem{pcm2010}
H.-S.~P. Wong, S.~Raoux, S.~Kim, J.~Liang, J.~P. Reifenberg, B.~Rajendran, M.~Asheghi, and K.~E. Goodson, ``Phase change memory,'' \emph{Proceedings of the IEEE}, vol.~98, no.~12, pp. 2201--2227, 2010.

\bibitem{Wu2019}
\BIBentryALTinterwordspacing
K.~Wu, G.~Dai, X.~Hu, S.~Li, X.~Xie, Y.~Wang, and Y.~Xie, ``\BIBforeignlanguage{en-US}{Memory-bound proof-of-work acceleration for blockchain applications},'' in \emph{\BIBforeignlanguage{en-US}{Proceedings of the 56th Annual Design Automation Conference 2019}}, Jun 2019. [Online]. Available: \url{http://dx.doi.org/10.1145/3316781.3317862}
\BIBentrySTDinterwordspacing

\bibitem{Xie2021}
X.~Xie, Z.~Liang, P.~Gu, A.~Basak, L.~Deng, L.~Liang, X.~Hu, and Y.~Xie, ``Spacea: Sparse matrix vector multiplication on processing-in-memory accelerator,'' in \emph{2021 IEEE International Symposium on High-Performance Computer Architecture (HPCA)}, 2021, pp. 570--583.

\bibitem{sim2015}
L.~Yavits, A.~Morad, and R.~Ginosar, ``Computer architecture with associative processor replacing last-level cache and {SIMD} accelerator,'' \emph{IEEE Transactions on Computers}, vol.~64, no.~2, pp. 368--381, 2015.

\bibitem{gamma2021}
\BIBentryALTinterwordspacing
G.~Zhang, N.~Attaluri, J.~S. Emer, and D.~Sanchez, ``Gamma: Leveraging gustavson’s algorithm to accelerate sparse matrix multiplication,'' in \emph{Proceedings of the 26th ACM International Conference on Architectural Support for Programming Languages and Operating Systems}, ser. ASPLOS '21.\hskip 1em plus 0.5em minus 0.4em\relax New York, NY, USA: Association for Computing Machinery, 2021, p. 687–701. [Online]. Available: \url{https://doi.org/10.1145/3445814.3446702}
\BIBentrySTDinterwordspacing

\bibitem{Zhang2018}
\BIBentryALTinterwordspacing
M.~Zhang, Y.~Zhuo, C.~Wang, M.~Gao, Y.~Wu, K.~Chen, C.~Kozyrakis, and X.~Qian, ``\BIBforeignlanguage{en-US}{Graphp: Reducing communication for pim-based graph processing with efficient data partition},'' in \emph{\BIBforeignlanguage{en-US}{2018 IEEE International Symposium on High Performance Computer Architecture (HPCA)}}, Feb 2018. [Online]. Available: \url{http://dx.doi.org/10.1109/hpca.2018.00053}
\BIBentrySTDinterwordspacing

\bibitem{Zhang2020}
Z.~Zhang, H.~Wang, S.~Han, and W.~J. Dally, ``Sparch: Efficient architecture for sparse matrix multiplication,'' in \emph{2020 IEEE International Symposium on High Performance Computer Architecture (HPCA)}, 2020, pp. 261--274.

\bibitem{Zheng2020}
L.~Zheng, J.~Zhao, Y.~Huang, Q.~Wang, Z.~Zeng, J.~Xue, X.~Liao, and H.~Jin, ``Spara: An energy-efficient reram-based accelerator for sparse graph analytics applications,'' in \emph{2020 IEEE International Parallel and Distributed Processing Symposium (IPDPS)}, 2020, pp. 696--707.

\bibitem{Zhou2022}
M.~Zhou, W.~Xu, J.~Kang, and T.~Rosing, ``Transpim: A memory-based acceleration via software-hardware co-design for transformer,'' in \emph{Proceedings of 2022 IEEE International Symposium on High-Performance Computer Architecture (HPCA)}, 2022, pp. 1071--1085.

\end{thebibliography}

\end{document}